\documentclass[journal=jpclcd,manuscript=article]{achemso}
\setkeys{acs}{maxauthors=10,etalmode=truncate,chaptertitle =true,articletitle=true}

\usepackage[hidelinks]{hyperref}
\usepackage{amssymb}
\usepackage{amsmath}
\usepackage[usenames,dvipsnames]{xcolor}
\usepackage{graphicx}
\usepackage{float}
\usepackage[normalem]{ulem}
\usepackage{caption}
\usepackage{subcaption}
\usepackage{natbib}
\newcommand{\bra}[1]{\left\langle #1 \right|}
\newcommand{\ket}[1]{\left|#1\right\rangle}
\newcommand{\braket}[2]{\left\langle#1 |  #2\right\rangle}

\usepackage{xcolor}

\usepackage{multirow}
\usepackage{bm}
\usepackage[version=4]{mhchem} 

\title{Relativistic Orbital Optimized Density Functional Theory for Accurate Core-Level Spectroscopy}

\author{Leonardo A. Cunha}
\email{leonardo.cunha@berkeley.edu}
\altaffiliation{These authors contributed equally to this work.}
\affiliation
{{Kenneth S. Pitzer Center for Theoretical Chemistry, Department of Chemistry, University of California, Berkeley, California 94720, USA}}
\alsoaffiliation{Chemical Sciences Division, Lawrence Berkeley National Laboratory, Berkeley, California 94720, USA}
\author{Diptarka Hait}
\email{diptarka@berkeley.edu}
\altaffiliation{These authors contributed equally to this work.}
\affiliation
{{Kenneth S. Pitzer Center for Theoretical Chemistry, Department of Chemistry, University of California, Berkeley, California 94720, USA}}
\alsoaffiliation{Chemical Sciences Division, Lawrence Berkeley National Laboratory, Berkeley, California 94720, USA}
\author{Richard Kang}
\affiliation
{{Kenneth S. Pitzer Center for Theoretical Chemistry, Department of Chemistry, University of California, Berkeley, California 94720, USA}}
\alsoaffiliation{Chemical Sciences Division, Lawrence Berkeley National Laboratory, Berkeley, California 94720, USA}
\author{Yuezhi Mao}
\affiliation{{Department of Chemistry, Stanford University, Stanford, CA 94305, USA}}
\author{Martin Head-Gordon}
\email{mhg@cchem.berkeley.edu}
\affiliation
{{Kenneth S. Pitzer Center for Theoretical Chemistry, Department of Chemistry, University of California, Berkeley, California 94720, USA}}
\alsoaffiliation{Chemical Sciences Division, Lawrence Berkeley National Laboratory, Berkeley, California 94720, USA}

\begin{document}

\maketitle
\begin{tocentry}
    \centering
	\includegraphics[scale = 1.0]{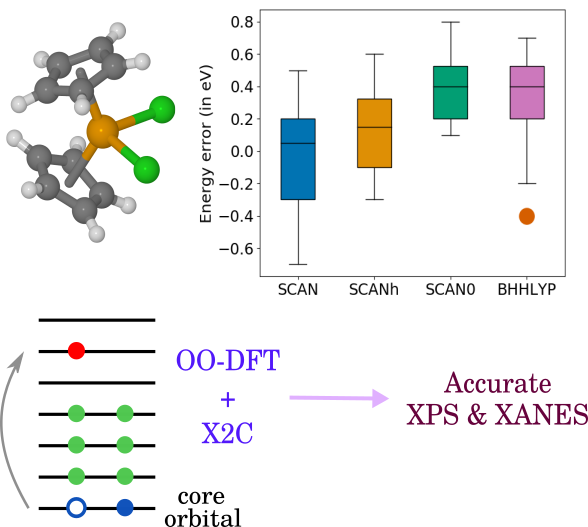}
\end{tocentry}

\begin{abstract}
Core-level spectra of 1s electrons of elements heavier than Ne show significant relativistic effects. We combine advances in orbital optimized DFT (OO-DFT) with the spin-free exact two-component (X2C) model for scalar relativistic effects, to study K-edge spectra of third period elements. OO-DFT/X2C is found to be quite accurate at predicting energies, yielding $\sim 0.5$ eV RMS error vs experiment with the modern SCAN (and related) functionals. This marks a significant improvement over the $>50$ eV deviations that are typical for the popular time-dependent DFT (TDDFT) approach. Consequently, experimental spectra are quite well reproduced by OO-DFT/X2C, sans empirical shifts for alignment. OO-DFT/X2C combines high accuracy with ground state DFT cost and is thus a promising route for computing core-level spectra of third period elements. We also explored K and L edges of 3d transition metals to identify limitations of the OO-DFT/X2C approach in modeling the spectra of heavier atoms.
\end{abstract}

Spectroscopy of core-level electrons with X-rays is a convenient and popular tool for studying chemical systems. A specific core-level of a given element normally has a characteristic energy that is quite distinct from inner-shells of other elements, making the technique element specific. Furthermore, core electrons do not play a direct role in chemical bonding, and are thus effectively localized around the nucleus. Spectroscopic probe of these electrons therefore yields information about local chemical environment of individual atoms. Core-level spectra can thus yield useful information about the local coordination environment\cite{yuhas2007probing,pollock2015insights}, extent of covalency in ligand-metal interactions\cite{westre1997multiplet,solomon2005ligand} or the oxidation state\cite{kubin2018probing}. Time-resolved core-level spectroscopy can also be used as a probe to study photoinduced chemical dynamics \cite{chergui2017photoinduced,bhattacherjee2018ultrafast,kraus2018ultrafast}.

Computational simulations of core-level spectra are useful for gaining insight into experiment, and potentially identifying new species whose signature may appear in transient spectra\cite{ochmann2017light,kraus2018ultrafast,bhattacherjee2017ultrafast}. Traditional quantum chemistry methods for excited states\cite{dreuw2005single,krylov2008equation} are however quite challenged by this task, especially since such techniques are mostly developed for (and validated on) problems involving only valence electrons. For example, the widely used linear-response time-dependent density functional theory (TDDFT) approach\cite{runge1984density,dreuw2005single} cannot adequately describe the relaxation of the core hole. This leads to rather large errors of $\sim$ 10-20 eV for the K-edge (1s orbitals) of C, N, O and F\cite{bhattacherjee2017ultrafast,wenzel2014calculating,lopata2012linear,besley2021modeling,zhang2012core,besley2020density}, if highly specialized functionals\cite{besley2009time,besley2010time} are not employed. Heavier elements lead to even larger errors, such as $\sim$ 50 eV for the P, S, Cl K-edges\cite{blake2018solid,martin2007sulfur,minasian2012determining} and $>100$ eV for the Fe K-edge\cite{debeer2008prediction}. TDDFT spectra therefore usually need to be empirically translated by many eVs, in order to align with experiment\cite{bhattacherjee2017ultrafast,attar2017femtosecond,debeer2008prediction,martin2007sulfur}. Similar behavior is observed for the equation-of-motion coupled cluster singles and doubles (EOM-CCSD) method\cite{stanton1993equation,krylov2008equation}, although the shift required is typically much smaller ($<2$ eV for second period elements) \cite{coriani2015communication,peng2015energy,frati2019coupled,vidal2019new,carbone2019analysis}. EOM-CCSD is however quite computationally demanding, with the computational cost scaling as $O(N^6)$ vs system size $N$ (compared to $O(N^{3-4})$ for DFT).

Orbital-optimized (OO) methods optimize orbitals for each excited state individually, and separately from those of the ground state. OO can therefore effectively model the relaxation of the core hole, leading to much better agreement with experiment\cite{besley2009self,hait2021orbital,derricotte2015simulation,zheng2020hetero} without any need for empirical shifts. Unfortunately, OO methods had been historically underutilized due to a risk of `variational collapse', in which the calculation converges to a lower energy state (often the ground state) instead of the desired high energy excitation.  However, there has been considerable recent interest in excited state OO, resulting in many new approaches that aim to reliably converge to any chosen state without the risk of variational collapse\cite{barca2018simple,shea2020generalized,ye2017sigma,hait2020excited,carter2020state,levi2020variational,grofe2020generalization}. \textcolor{black}{In practice, OO-DFT methods require more compute time than TDDFT if a large number of states are desired, such as in (near-)degenerate bands. This stems from OO-DFT having to iteratively optimize multiple states individually while TDDFT can simultaneously compute them. However, OO-DFT retains the same computational scaling as ground state DFT or TDDFT. An overview of the successes and challenges with OO-DFT methods can be found in Ref \citenum{hait2021orbital}}. OO-DFT methods are thus increasingly being employed to study core-level spectra\cite{hait2020highly,hait2020accurate,carter2020state,garner2020core,zhao2021dynamic,kahk2021core}, with the modern SCAN\cite{SCAN} functional leading to very low error\cite{kahk2019accurate,hait2020highly,hait2020accurate,kahk2021core} ($<1$ eV) vs experiment for the K-edge of C, N, O and F, as well as L-edges of Si, P, S and Cl. 

The K-edge of elements heavier than F however cannot be as accurately modeled with non-relativistic quantum mechanics. Naive use of the Bohr atom model suggests that the speed of 1s electrons would scale linearly with the atomic number $Z$, eventually attaining the speed of light at $Z > \alpha^{-1} \approx 137$ (where $\alpha$ is the fine structure constant). Relativistic effects become perceptible at much smaller $Z$, with calculations indicating that non-relativistic quantum mechanics underbinds the 1s electrons of Ne by 1 eV\cite{takahashi2017relativistic}. It is therefore necessary to incorporate relativistic effects into OO-DFT, if $<1$ eV error vs experiment is desired for computed K-edge spectra of third period elements and beyond. \textcolor{black}{Scalar relativistic treatment is however often overlooked for linear-response TDDFT, as the ad-hoc empirical shifts (typically larger than the relativistic correction) utilized to align computation with experiment account for it to some extent\cite{norman2018simulating,bussy2021efficient,stetina2019modeling}. Nonetheless, explicit use of relativistic effects in TDDFT has been previously explored, albeit mostly within a real-time framework\cite{repisky2015excitation,lopata2012linear,stetina2019modeling}. Similarly, both empirical shifts \cite{peng2015energy} and explicit inclusion\cite{liu2021relativistic,halbert2021relativistic} have been used to account for relativistic effects in coupled cluster methods. The use of element-specific corrections to nonrelativistic TDDFT results has also been examined in the past\cite{besley2009time}.}

In this work, we utilize the spin-free exact-two component one electron (SFX2C-1e, henceforth referred to as just X2C) model for relativistic quantum chemistry\cite{dyall1997interfacing,kutzelnigg2005quasirelativistic,saue2007inf, Liu2009x2c,saue2011primer, Liu2012spin,cheng2011analytic,verma2016predicting} to obtain improved OO-DFT core-level spectra. 
The X2C model transforms the one particle terms of the electronic Hamiltonian (i.e. kinetic energy and external potential) via the solutions of the four component, one electron Dirac Hamiltonian. The two-particle (i.e. interelectron interaction) terms are treated within the pure Coulomb formalism and, therefore, are left unaltered in the non-relativistic form, permitting straightforward application of DFT. The transformation is briefly described in the supporting information, and we invite interested readers to examine Refs. \citenum{saue2011primer} and \citenum{Liu2012spin} for further details regarding X2C. 

Using the X2C transformed one particle Hamiltonian, we obtained the ground state energy via the standard Kohn-Sham\cite{kohn1965self} (KS) formalism. Excited states are more challenging, as many excitations unpair electrons and therefore require multiple Slater determinants for a spin-pure description. DFT for such states is not straightforward, as the KS-DFT formalism with existing density functional approximations can only be reliably applied to single-reference systems. 
We therefore utilize three related OO-DFT ansatze for modeling three classes of excited states. These ansatze are described in detail in Ref \citenum{hait2021orbital}, but we provide a brief outline here for convenience. 

\begin{figure}[htb!]
    \centering
\begin{minipage}{0.5336\textwidth}
    \centering
    \includegraphics[width=\linewidth]{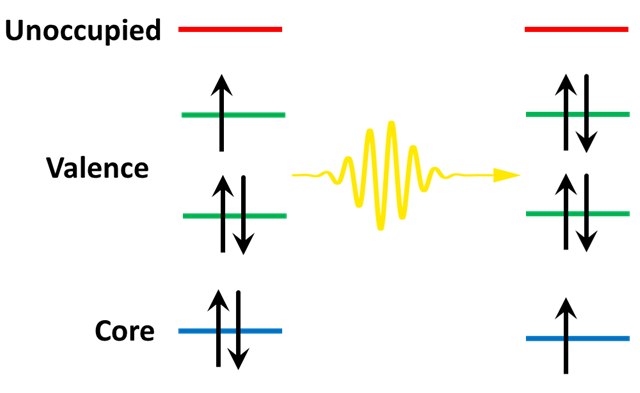}
    \subcaption{Core$\to$SOMO excitation in a radical.}
    \label{fig:dscfradical}
\end{minipage}
\hfill
\begin{minipage}{0.4264\textwidth}
    \centering
    \includegraphics[width=\linewidth]{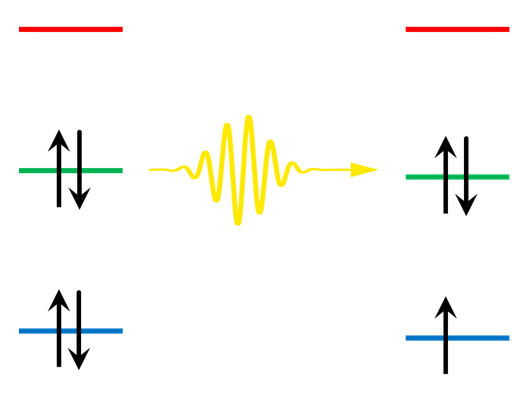}
    \subcaption{Core-ionization of closed-shell system}
    \label{fig:dscfion}
\end{minipage}
\caption{Schematic for processes where $\Delta$SCF is appropriate.
}
\label{fig:dscf}
\end{figure}

\begin{figure}[htb!]
    \centering
    \includegraphics[width=0.8\linewidth]{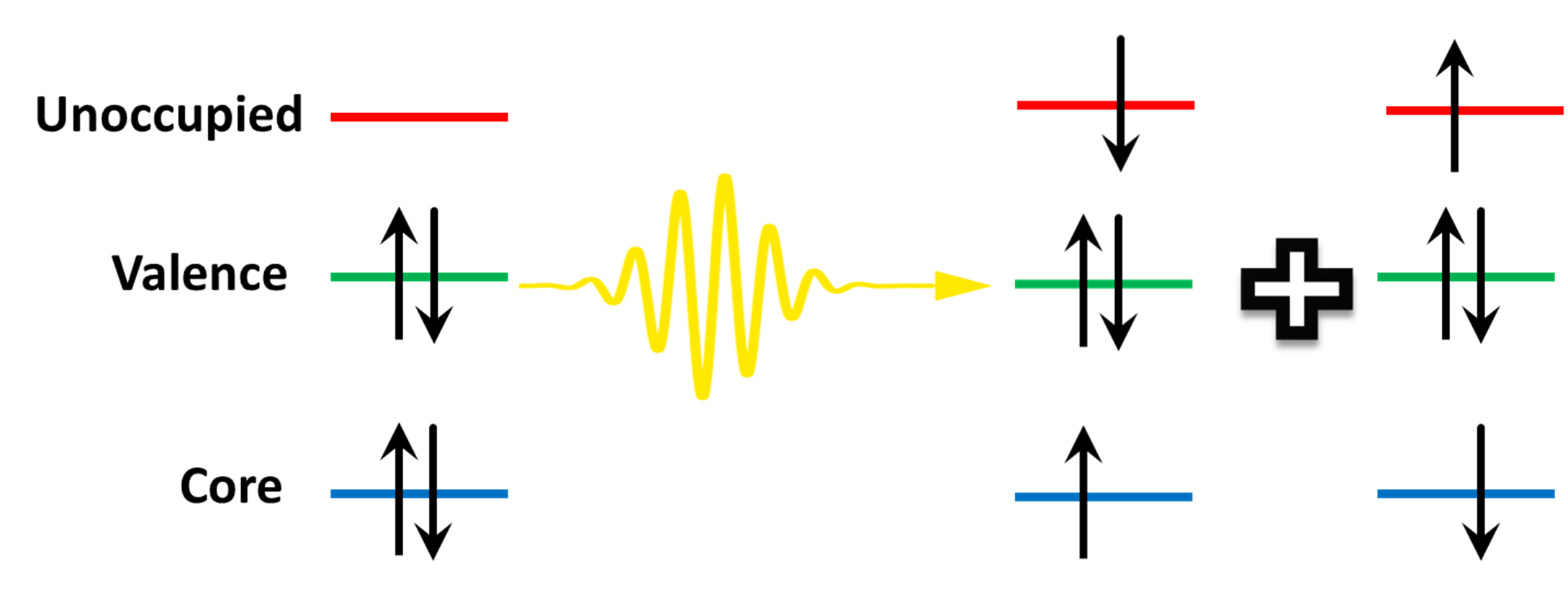}
\caption{Schematic for a singlet core excitation in a closed-shell species. Two open-shell determinants (as seen on the right) are equally likely, and are individually halfway between singlet and triplet in character. ROKS is thus essential for spin-purity of the excited state. 
}
\label{fig:roks}
\end{figure}

A state with no unpaired electrons, or one with all unpaired electrons of the same spin can be represented by a single Slater determinant. It is straightforward to optimize a single Slater determinant with an excited state electronic configuration (apart from the aforementioned risk of variational collapse). This protocol is called $\Delta$SCF\cite{bagus1965self,ziegler1977calculation} and is suitable for states that result from 1s$\to$ SOMO (singly occupied molecular orbital) transitions of open-shell species, or core-ionized states of closed-shell molecules (as shown in Fig \ref{fig:dscf}). \textcolor{black}{Relativistic $\Delta$ Hartree-Fock (HF) has indeed been used to study core-ionization energies\cite{niskanen2011relativistic,zheng2019performance}}. However, $\Delta$SCF is not appropriate for singly excited singlet excited states of closed-shell molecules, as both the up and down spins are equally likely to be excited (as shown in Fig \ref{fig:roks}). Exciting only one spin results in a spin-contaminated determinant midway between singlet and triplet. \textcolor{black}{Spin-contaminated $\Delta$SCF energies have nonetheless been utilized in the past for core-excitation energy calculations, with element-specific relativistic corrections for heavy elements\cite{besley2009self}.}

Restricted open-shell Kohn-Sham\cite{frank1998molecular,kowalczyk2013excitation} (ROKS) obtains a pure singlet energy by spin-projection on the spin-contaminated determinant. 
ROKS is consequently the optimal OO-DFT approach for singlet excited states with two unpaired spins, although it cannot be applied if there are more than two unpaired spins. Such states require a more general recoupling scheme described in Refs \citenum{hait2021orbital} and \citenum{hait2020accurate}. This is however only necessary for transitions from the core to completely unoccupied levels in open-shell systems, with $\Delta$SCF and ROKS being sufficient for all the states considered in this work. \textcolor{black}{OO-DFT therefore encompasses $\Delta$SCF and methods like ROKS and the general recoupling scheme that derive from it.}

\begin{table}[htb!]
\begin{tabular}{lrrrrrrr}\hline\hline
\multicolumn{1}{c}{Molecule} & \multicolumn{1}{c}{Expt.} & \multicolumn{1}{c}{SCAN (NR)} & \multicolumn{1}{c}{SCAN} & \multicolumn{1}{c}{SCANh} & \multicolumn{1}{c}{SCAN0} & \multicolumn{1}{c}{BHHLYP} & \multicolumn{1}{c}{HF} \\\hline
Ne                           & 870.2\cite{aagren1978multiplet}                     & 869.3                       & 870.3                    & 870.3                     & 870.3                     & 870.5                      & 869.6                  \\
Mg                           & 1311.5\cite{banna1978free}                    & 1309.3                      & 1311.6                   & 1311.6                    & 1311.6                    & 1311.5                     & 1311.4                 \\
\ce{SiH4}                        & 1847.0\cite{bodeur1990single}                     & 1842.7                      & 1847.1                   & 1847.2                    & 1847.3                    & 1847.3                     & 1848.0                 \\
\ce{SiF4}                        & 1852.5\cite{bodeur1990single}                    & 1847.8                      & 1852.1                   & 1852.4                    & 1852.7                    & 1852.8                     & 1853.7                 \\
\ce{\textbf{Si}Cl4}                          & 1850.6\cite{bodeur1990single}                     & 1846.1                      & 1850.5                   & 1850.7                    & 1850.9                    & 1851.0                     & 1852.0                 \\
\ce{SiBr4}                           & 1849.7\cite{bodeur1990single}                    & 1845.5                      & 1849.8                   & 1850.0                    & 1850.3                    & 1850.3                     & 1851.3                 \\
\ce{PH3}                          & 2150.9\cite{sodhi1983kll}                    & 2145.1                      & 2151.0                   & 2151.1                    & 2151.2                    & 2151.0                     & 2152.0                 \\
\ce{PF3}                           & 2156.4\cite{sodhi1983kll}                    & 2150.0                      & 2155.8                   & 2156.1                    & 2156.4                    & 2156.4                     & 2157.7                 \\
\ce{PF5}                           & 2159.4\cite{sodhi1983kll}                    & 2153.2                      & 2159.0                   & 2159.3                    & 2159.8                    & 2160.0                     & 2161.6                 \\
\ce{POF3}                          & 2157.8\cite{sodhi1983kll}                    & 2151.7                      & 2157.5                   & 2157.8                    & 2158.2                    & 2158.4                     & 2159.9                 \\
\ce{H2S}                          & 2478.5\cite{keski1976energies}                   & 2470.7                      & 2478.4                   & 2478.6                    & 2478.8                    & 2478.4                     & 2479.5                 \\
\ce{CS2}                          & 2478.1\cite{perera1984molecular}                    & 2470.4                      & 2478.2                   & 2478.3                    & 2478.5                    & 2478.0                     & 2479.1                 \\
\ce{SF4}                          & 2486.9\cite{sodhi1986kll}                    & 2478.7                      & 2486.5                   & 2486.8                    & 2487.3                    & 2487.4                     & 2489.3                 \\
\ce{SF6}                          & 2490.1\cite{keski1976energies}                    & 2481.9                      & 2489.6                   & 2489.9                    & 2490.5                    & 2490.6                     & 2492.7                 \\
\ce{SO2}                          & 2483.7\cite{keski1976energies}                    & 2475.9                      & 2483.6                   & 2483.9                    & 2484.3                    & 2484.2                     & 2486.0                 \\
CSO                          & 2478.7\cite{perera1984molecular}                    & 2471.2                      & 2479.0                   & 2479.1                    & 2479.3                    & 2478.8                     & 2479.9                 \\
\ce{\textbf{S}F5Cl}             & 2488.9\cite{sodhi1986kll}                    & 2480.9                      & 2488.6                   & 2489.0                    & 2489.5                    & 2489.5                     & 2491.5                 \\
HCl                          & 2829.8\cite{bodeur1990chlorine}                    & 2820.3                      & 2830.3                   & 2830.4                    & 2830.6                    & 2830.0                     & 2831.4                 \\
\ce{Cl2}                         & 2830.2\cite{bodeur1990chlorine}                     & 2820.8                      & 2830.8                   & 2831.0                    & 2831.2                    & 2830.5                     & 2831.9                 \\
\ce{CH3Cl}                        & 2828.4\cite{lindle1991polarized}                    & 2819.2                      & 2829.2                   & 2829.3                    & 2829.5                    & 2828.9                     & 2830.2                 \\
\ce{SF5\textbf{Cl}}            & 2829.6\cite{reynaud1992electronic}                   & 2820.3                      & 2830.4                   & 2830.6                    & 2830.9                    & 2830.3                     & 2831.8                 \\
\ce{CCl3F}                        & 2829.3\cite{lindle1991polarized}                    & 2820.0                      & 2830.0                   & 2830.2                    & 2830.4                    & 2829.8                     & 2831.2                 \\
Ar                           & 3206.3\cite{breinig1980atomic}                    & 3194.1                      & 3206.9                   & 3207.0                    & 3207.3                    & 3206.5                     & 3208.1                 \\
\hline
\multicolumn{1}{l}{RMSE}     & \multicolumn{1}{l}{}      & 7.4                         & 0.4                      & 0.5                       & 0.6                       & 0.4                        & 1.7                    \\
\multicolumn{1}{l}{ME}       & \multicolumn{1}{l}{}      & -6.9                        & 0.1                      & 0.3                       & 0.5                       & 0.3                        & 1.5                    \\
\multicolumn{1}{l}{MAX}      & \multicolumn{1}{l}{}      & 12.2                        & 0.8                      & 1.0                       & 1.3                       & 0.7                        & 2.7                   \\ \hline\hline
\end{tabular}
\caption{Gas phase XPS K-edge binding energies for Ne and third period elements (in eV). Computed values were found from restricted open-shell $\Delta$SCF calculations, using the aug-pcX-2 basis\cite{ambroise2018probing} when available and decontracted aug-pcseg-2\cite{jensen2014unifying} for H/Br. Non-relativistic (NR) values from SCAN are also provided for comparison. The root mean square error (RMSE), mean error (ME) and maximum absolute error (MAX) are also reported. The atomic site of the ionization is \textbf{bolded} when multiple possibilities exist.}
\label{tab:gasxps}
\end{table}


We first examined the performance of OO-DFT/X2C in predicting the gas phase K-edge spectra of the third period elements (and Ne) with Table \ref{tab:gasxps} reporting  $\Delta$SCF 1s electron binding energies for several closed-shell species. All presented functionals have root mean square error (RMSE) $<1$ eV vs experimental X-ray photoelectron spectra (XPS). These functionals were identified via screening across many functionals over a smaller set of species (\ce{SiH4},\ce{PH3},\ce{H2S},\ce{HCl} and Ar). This screening also revealed that other well known functionals like B3LYP\cite{b3lyp,stephens1994ab}, PBE0\cite{pbe0}, or TPSS\cite{tpss} have larger errors ($\sim$ 1-3  eV, as shown in the supporting information) that nonetheless represent a major improvement over TDDFT or non-relativistic OO-DFT. Out of the selected functionals, SCAN fares particularly well, yielding an RMSE of 0.4 eV and a maximum deviation of 0.8 eV from experiment. X2C is crucial for this level of agreement, as SCAN with the non-relativistic (NR) Hamiltonian leads to errors of several eV (as shown in Table \ref{tab:gasxps}). The related SCANh\cite{chan2021assessment} functional performs slightly worse, but is still fairly accurate. SCANh does have positive mean error (ME), indicating it systematically overestimates the binding energy. This overestimation is a consequence of the presence of HF exchange (10\%) in the functional, as pure HF overestimates by $\sim 2$ eV. Overestimation is more evident for SCAN0\cite{scan0} (which has 25\% HF exchange) and BHHLYP\cite{bhhlyp} (50\% HF exchange). However, it is important to note that the ME is strongly influenced by the choice of the local exchange-correlation model. For example, functionals based on PBE\cite{PBE} appear to be far more sensitive to \% HF exchange, than ones derived from SCAN (as seen in the supporting information).  In addition, most local functionals strongly underbind core-electrons, and would require admixture of a very large amount of \% HF exchange to have low error (BHHLYP being a prominent example). SCAN is a notable exception in this regard, as it has low error despite being a local functional.

\begin{table}[htb!]
\begin{tabular}{lrrrrrrr}\hline\hline
\multicolumn{1}{c}{Molecule} & \multicolumn{1}{c}{Expt.} & \multicolumn{1}{c}{SCAN (NR)} & \multicolumn{1}{c}{SCAN} & \multicolumn{1}{c}{SCANh} & \multicolumn{1}{c}{SCAN0} & \multicolumn{1}{c}{BHHLYP} & \multicolumn{1}{c}{HF} \\\hline 
\ce{SiH4}                     & 1842.7\cite{bodeur1990single}               & 1838.56                  & 1842.9                   & 1843.1                  & 1843.3                  & 1843.4                  & 1844.9                  \\
\ce{SiF4}                     & 1849.0\cite{bodeur1990single}               & 1844.13                  & 1848.5                   & 1848.8                  & 1849.1                  & 1849.4                  & 1851.3                  \\
\ce{\textbf{Si}Cl4}                    & 1846.0\cite{bodeur1990single}               & 1841.37                  & 1845.7                   & 1845.9                  & 1846.2                  & 1846.5                  & 1848.5                  \\
\ce{SiBr4}                    & 1845.0\cite{bodeur1990single}               & 1840.58                  & 1844.9                   & 1845.1                  & 1845.4                  & 1845.6                  & 1847.7                  \\
\ce{PH3}                      & 2145.8\cite{cavell1999chemical}               & 2140.15                  & 2146.0                   & 2146.2                  & 2146.4                  & 2146.3                  & 2148.1                  \\
\ce{PF3}                      & 2149.3\cite{cavell1999chemical}               & 2143.55                  & 2149.4                   & 2149.6                  & 2149.9                  & 2150.0                  & 2151.9                  \\
\ce{PF5}                      & 2155.0\cite{cavell1999chemical}               & 2148.59                  & 2154.5                   & 2154.7                  & 2155.2                  & 2155.5                  & 2159.8                  \\
\ce{POF3}                     & 2153.3\cite{cavell1999chemical}               & 2147.16                  & 2153.0                   & 2153.3                  & 2153.7                  & 2153.9                  & 2158.0                  \\
\ce{H2S}                      & 2472.7\cite{reynaud1996double}               & 2465.05                  & 2472.8                   & 2473.0                  & 2473.2                  & 2472.9                  & 2475.0                  \\
\ce{CS2}                      & 2470.8\cite{perera1984molecular}               & 2463.53                  & 2471.3                   & 2471.3                  & 2471.4                  & 2471.0                  & 2472.5                  \\
\ce{SF4}                      & 2477.3\cite{bodeur1987inner}               & 2469.71                  & 2477.4                   & 2477.7                  & 2478.0                  & 2478.0                  & 2480.5                  \\
\ce{SF6}                      & 2486.0\cite{reynaud1996double}               & 2477.55                  & 2485.3                   & 2485.7                  & 2486.2                  & 2486.5                  & 2490.1                  \\
\ce{SO2}                      & 2473.2\cite{reynaud1996double}               & 2465.53                  & 2473.3                   & 2473.4                  & 2473.6                  & 2473.4                  & 2475.5                  \\
CSO                      & 2472.0\cite{perera1984molecular}               & 2464.73                  & 2472.5                   & 2472.6                  & 2472.7                  & 2472.2                  & 2473.9                  \\
\ce{\textbf{S}F5Cl}         & 2483.5\cite{reynaud1992electronic}               & 2475.11                  & 2482.8                   & 2483.2                  & 2483.7                  & 2483.9                  & 2487.3                  \\
HCl                      & 2823.9\cite{bodeur1990chlorine}               & 2813.79                  & 2823.8                   & 2824.0                  & 2824.2                  & 2823.7                  & 2825.7                  \\
\ce{Cl2}                      & 2821.3\cite{bodeur1990chlorine}               & 2811.05                  & 2821.1                   & 2821.2                  & 2821.4                  & 2820.9                  & 2822.8                  \\
\ce{CH3Cl}                    & 2823.5\cite{lindle1991polarized}               & 2813.60                  & 2823.6                   & 2823.8                  & 2824.0                  & 2823.6                  & 2825.8                  \\
\ce{SF5\textbf{Cl}}        & 2821.8\cite{reynaud1992electronic}               & 2811.83                  & 2821.8                   & 2821.9                  & 2822.1                  & 2821.6                  & 2823.7                  \\
\ce{CCl3F}                    & 2822.8\cite{lindle1991polarized}               & 2813.19                  & 2823.2                   & 2823.4                  & 2823.6                  & 2823.2                  & 2825.3                  \\\hline 
\multicolumn{1}{l}{RMSE} & \multicolumn{1}{l}{} & \multicolumn{1}{r}{7.6}  & \multicolumn{1}{r}{0.4}  & \multicolumn{1}{r}{0.3} & \multicolumn{1}{r}{0.5} & \multicolumn{1}{r}{0.5} & \multicolumn{1}{r}{2.8} \\
\multicolumn{1}{l}{ME}   & \multicolumn{1}{l}{} & \multicolumn{1}{r}{-7.3} & \multicolumn{1}{r}{-0.1} & \multicolumn{1}{r}{0.1} & \multicolumn{1}{r}{0.4} & \multicolumn{1}{r}{0.3} & \multicolumn{1}{r}{2.7} \\
\multicolumn{1}{l}{MAX}  & \multicolumn{1}{l}{} & \multicolumn{1}{r}{10.3} & \multicolumn{1}{r}{0.7}  & \multicolumn{1}{r}{0.6} & \multicolumn{1}{r}{0.8} & \multicolumn{1}{r}{0.7} & \multicolumn{1}{r}{4.8}\\\hline\hline
\end{tabular}
\caption{Lowest dipole allowed gas-phase XAS excitation energy for third period elements (in eV). Computed values were found from ROKS, using the aug-pcX-2 basis when available and decontracted aug-pcseg-2 for H/Br. NR values from SCAN are also provided for comparison. The atomic site of the ionization is \textbf{bolded} when multiple possibilities exist.}
\label{tab:gasxas}
\end{table}

We next considered prediction of X-ray absorption spectra (XAS) with ROKS, which is quite effective in predicting singlet core excitation energies of second period elements\cite{hait2020highly}. Table  \ref{tab:gasxas} shows that inclusion of scalar relativistic effects through the X2C model permits high accuracy for third period elements as well. SCANh yields the best performance with an RMSE of 0.3 eV and a maximum absolute error (MAX) of only 0.6 eV. SCAN and SCAN0 also yield quite good performance. In fact, the RMSE for all the presented functionals is comparable to the typical experimental energy resolution of $\sim 0.5$ eV, and therefore indicative of semi-quantitative performance. Curiously, SCAN significantly underestimates the excitation energy for highly fluorinated compounds (SF$_6$, CF$_3$SF$_5$ etc.), highlighting a potential limitation for this otherwise excellent performing local functional. This systematic underestimation is partially mitigated with HF exchange, leading to SCANh performing somewhat better. On the other hand, SCAN0 has a systematic bias towards overestimation due to a greater part of HF exchange being present. SCANh therefore offers a reasonable middle path, although it would perform poorly for cases where SCAN already overestimates or SCAN0 underestimates.  We also note that our RMSEs are considerably smaller than the several eV errors reported by an earlier study\cite{verma2016predicting} using \textcolor{black}{relativistic} orthogonality constrained DFT (OC-DFT)\cite{evangelista2013orthogonality}, which may in part stem from use of B3LYP in that work.  

\begin{figure}[htb!]
    \centering
\begin{minipage}{0.48\textwidth}
    \centering
    \includegraphics[width=\linewidth]{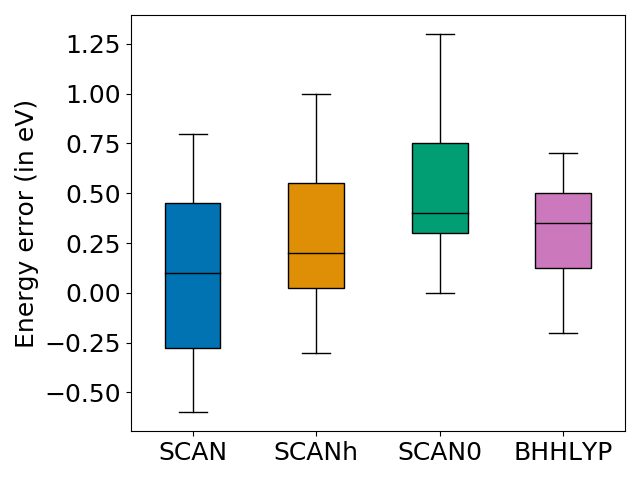}
    \subcaption{K-edge ionization.}
    \label{fig:xpsbox}
\end{minipage}\hfill
\begin{minipage}{0.48\textwidth}
    \centering
    \includegraphics[width=\linewidth]{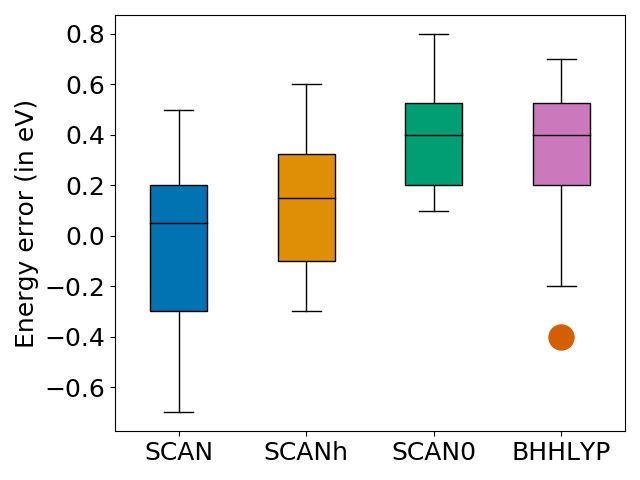}
    \subcaption{Lowest dipole allowed K-edge excitation.}
    \label{fig:xasbox}
\end{minipage}
\caption{Box plot for errors (vs experiment) in computed values reported in Tables \ref{tab:gasxps}-\ref{tab:gasxas}. The red dot in Fig. \ref{fig:xasbox} represents an outlier (\ce{Cl2}) where BHHLYP unusually underestimates the excitation energy by 0.4 eV.}
\label{fig:boxplots}
\end{figure}

Fig  \ref{fig:boxplots} visually summarizes the key result shown in Tables \ref{tab:gasxps} and \ref{tab:gasxas}, namely that OO-DFT/X2C is effective in predicting core-level excitation/ionization energies of third period atoms in isolated small molecules. The error distributions are quite compact overall, with the typical limits being below 1 eV. In particular, the SCAN, SCANh and BHHLYP functionals typically have errors below 0.5 eV, and never over 1.0 eV (for the species considered). They therefore appear to be promising routes for prediction of gas-phase core-level spectra.

It is also worthwhile to consider larger systems in order to gauge feasibility of OO-DFT/X2C for widespread practical use. However, certain computational challenges need to be considered along the way. Decontracted basis sets of at least triple-$\zeta$ quality (ideally of the Jensen pcX-n\cite{ambroise2018probing} or Dunning cc-pCVnZ\cite{woon1995gaussian} type) appear to be necessary for computation of core-level spectra, both to account for relaxation of the core hole and for convergence of relativistic effects. However, using such bases for all atoms would be quite computationally demanding. The local nature of the core-excitation permits use of a mixed basis strategy\cite{hait2020highly} in which the decontracted triple-$\zeta$ basis is only used for the atom whose core-electrons are being probed, while the corresponding contracted double-$\zeta$ basis is sufficient for all other atoms. We have verified that this mixed basis strategy does not lead to any significant change in RMSE of SCAN/SCANh for the species in Tables \ref{tab:gasxps} and \ref{tab:gasxas} (shown in the supporting information) and have employed this strategy for the calculations reported hereon. 

In addition, XAS for larger systems is often collected in the solid state or in solution, making it necessary to model the effect of the environment on the calculated spectrum. The locality of core-electrons suggests that only the first coordination shell needs to be considered atomistically for the first few (`pre edge') peaks, with a continuum dielectric model being adequate for the remainder of the environment. Such polarizable continuum models (PCM\cite{mennucci2012polarizable}) are likely to be effective for species in which the core electron is excited to a valence level, but would probably be insufficient for Rydberg-like excitations without atomistic modeling of a rather large region around the core-hole. We have employed the integral equation formalism (IEF-) PCM\cite{cances1997new} model to account for environment effects, and thereby investigate how spectra are affected by the phase of the system\cite{kunze2021pcm}. IEF-PCM is likely adequate if it only induces a small shift relative to vacuum results, but more sophisticated embedding techniques might prove necessary if there is a significant difference between IEF-PCM and vacuum calculations. 

\begin{table}[htb!]
\begin{tabular}{lllrr}\hline\hline
Molecule                                            & Environment & \multicolumn{1}{l}{Expt.} & \multicolumn{1}{l}{SCAN} & \multicolumn{1}{l}{SCANh} \\\hline
\ce{\textbf{Si}(Me)4}                                             & Gas         & 1843.6\cite{sutherland1993si}                    & 1843.5                   & 1843.7                    \\
\ce{\textbf{Si}(OMe)4}                                            & Gas         & 1845.9\cite{sutherland1993si}                    & 1845.9                   & 1846.1                    \\
\ce{(CH3O)2\textbf{P}(S)Cl}                                       & Gas         & 2150.2\cite{cavell1999chemical}                    & 2150.0                   & 2150.2                    \\
\ce{\textbf{P}4O6}                                                & Gas         & 2147.5\cite{engemann1997experimental}                    & 2147.8                   &   2148.0    \\
\ce{O\textbf{P}Ph3}                                                & Solid         & 2147.3\cite{blake2018solid}                    & 2147.4                   &  2147.7    \\
\ce{CF3\textbf{S}F5}                                              & Gas         & 2483.8\cite{ibuki2004total}                    & 2483.2                   & 2483.6                    \\
\ce{CH3\textbf{S}SCH3}                                              & Gas         & 2471.6\cite{ochmann2018uv}                    & 2472.0                   & 2472.1                    \\
\ce{(4-Me)C6H4\textbf{S}H}                                  & Cyclohexane & 2472.5\cite{ochmann2017light}                    & 2472.5                   & 2472.7                    \\
\ce{(4-Me)C6H4\textbf{S}}$\cdot$ & Cyclohexane & 2467.0\cite{ochmann2017light}                    & 2467.4                   & 2467.5                    \\
\ce{Ti\textbf{Cl}4}                                               & Toluene     & 2821.6\cite{debeer2005metal}                    & 2821.3                   & 2821.2    \\  \hline\hline             
\end{tabular}
\caption{Lowest dipole allowed XAS excitation energy for slightly larger species (in eV) from experiment and theory. A mixed basis (aug-pcX-2 on excitation site, aug-pcseg-1 on all other atoms) was utilized for the calculations.}
\label{tab:largegas}
\end{table}

Table \ref{tab:largegas} shows performance for the mixed basis protocol for molecular systems in the gas phase, or non-polar solvents like cyclohexane (modeled with IEF-PCM, if present). ROKS was used for all closed-shell systems, while single determinant spin-unrestricted $\Delta$SCF was sufficient\cite{hait2020accurate} for the S 1s $\to$ SOMO transition of the 4-methylthiophenoxy (\ce{(4-Me)C6H4\textbf{S}}) radical. Our approach appears to be quite accurate in predicting experimental energies, indicating that the OO-DFT/X2C approach can be applied to large molecules for prediction of heavy element K-edges. We also revisited earlier work on light elements\cite{hait2020accurate}, and demonstrated that inclusion of X2C does not cause any degradation of performance in predicting excitation energies (as shown in the supporting information). 

\begin{table}[htb!]
\begin{tabular}{lr|rrr|rrr}\hline\hline
Species     & \multicolumn{1}{c}{Experiment} & \multicolumn{3}{c}{Vacuum}                                                       & \multicolumn{3}{c}{Solid}                                                        \\\hline
            & \multicolumn{1}{c}{}           & \multicolumn{1}{c}{SCAN} & \multicolumn{1}{c}{SCANh} & \multicolumn{1}{c}{SCAN0} & \multicolumn{1}{c}{SCAN} & \multicolumn{1}{c}{SCANh} & \multicolumn{1}{c}{SCAN0} \\\hline
\ce{ClO4^-}        & 2835.1\cite{mckeown2011x}                         & 2834.6                   & 2834.9                    & 2835.4                    & 2834.7                   & 2835.0                    & 2835.4                    \\
\ce{ClO3^-}           & 2831.3\cite{mckeown2011x}                         & 2830.8                   & 2831.0                    & 2831.4                    & 2830.8                   & 2831.1                    & 2831.4                    \\
\ce{ClO2^-}           & 2826.8\cite{mckeown2011x}                         & 2826.5                   & 2826.7                    & 2827.0                    & 2826.6                   & 2826.8                    & 2827.1                    \\
            & \multicolumn{1}{l}{}           & \multicolumn{1}{l}{}     & \multicolumn{1}{l}{}      & \multicolumn{1}{l}{}      & \multicolumn{1}{l}{}     & \multicolumn{1}{l}{}      & \multicolumn{1}{l}{}      \\
\ce{CuCl4}$^{2-}$ & 2820.2\cite{shadle1995ligand}                         & 2820.1                   & 2820.3                    & 2820.6                    & 2819.8                   & 2819.9                    & 2820.2                    \\
\ce{NiCl4}$^{2-}$       & 2821.5\cite{shadle1995ligand}                         & 2821.4                   & 2821.6                    & 2822.0                    & 2821.1                   & 2821.2                    & 2821.6                    \\
\ce{CoCl4}$^{2-}$       & 2822.5\cite{shadle1995ligand}                         & 2822.3                   & 2822.0                    & 2822.5                    & 2821.4                   & 2822.4                    & 2822.1                    \\
\ce{FeCl4}$^{2-}$   & 2822.8\cite{shadle1995ligand}                         & 2822.0                   & 2822.3                    & 2822.7                    & 2821.7                   & 2821.9                    & 2822.3                   \\
\ce{FeCl4}$^{-}$  & 2820.5\cite{shadle1995ligand}                         & 2820.4                   & 2820.4                    & 2820.4                    & 2820.2                   & 2820.1                    & 2820.1 \\\hline 
RMSE & & 0.4  & 0.3  & 0.3 & 0.6  & 0.4  & 0.3  \\
 ME &   & -0.3 & -0.2 & 0.2 & -0.6 & -0.3 & -0.1 \\
MAX   & & 0.8  & 0.5  & 0.5 & 1.1  & 0.9  & 0.5  \\\hline\hline
\end{tabular}
\caption{Lowest dipole allowed Cl K-edge excitation energy for ionic species (in eV). All experimental data correspond to solid state measurements. ROKS was used for closed-shell species like \ce{ClO4^-}, and spin-unrestricted $\Delta$SCF for open-shell systems like \ce{CuCl4}$^{2-}$. A mixed basis set (aug-pcX-2 on excitation site, aug-pcseg-1 on all other atoms) was utilized.  }
\label{tab:clion}
\end{table}

Although we have only considered neutral species so far,  core-level spectra of ionic moieties have also been collected in many experiments. These species offer an interesting regime for both testing the efficacy of our approach and for gauging environment effects in general. Table \ref{tab:clion} presents a comparison between experiment and theory for Cl K-edges of several ionic species. The ions in Table \ref{tab:clion} can be broadly categorized into two categories. The first are closed-shell species where Cl has a formally positive oxidation state (ClO$_4^-$ etc) and the lowest excitation is 1s $\to\sigma^*_{\textrm{O-Cl}}$.  These $\sigma^*$ orbitals are more `Cl like' due to the halogen being electropositive, leading to excitations  that are thus mostly localized on the Cl (which is at the center of the ion) and therefore reasonably isolated from the environment. Consequently, not much difference is found between predictions for vacuum, and an IEF-PCM model ionic solid. SCAN, SCANh and SCAN0 all fare reasonably at predicting excitation energies, with the former slightly underestimating and the latter slightly overestimating. 

The second class of ions are high-spin tetrahedral transition metal chloride complexes where the lowest transition is charge-transfer (CT) from Cl to a singly occupied metal d level. In addition, the Cl site is on the periphery of the molecule, permitting greater influence from the environment. There is thus a perceptible red-shift in the IEF-PCM results relative to vacuum (due to greater stabilization of the CT like excited state). In addition, SCAN0 appears to be the best performer for this class of excitations. Nonetheless, SCANh appears to do a reasonable job at predicting excitation energies for all of the ionic systems, indicating that the OO-DFT/X2C approach remains capable of delivering semi-quantitative accuracy even outside of small molecules in the gas phase. 

\begin{figure}[htb!]
    \centering
\begin{minipage}{0.48\textwidth}
    \centering
    \includegraphics[width=\linewidth]{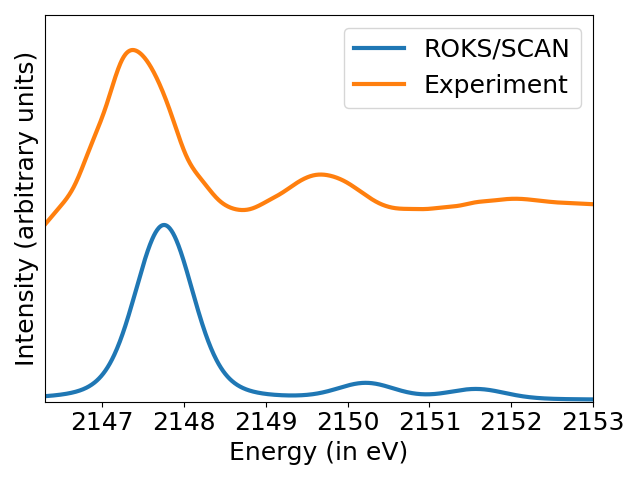}
    \subcaption{P K-edge XAS of gaseous \ce{P4O6}\cite{engemann1997experimental}, using a mixed basis (aug-pcX-2 on target P, aug-pcseg-1 on other atoms).}
    \label{fig:p4o6}
\end{minipage}\hfill
\begin{minipage}{0.48\textwidth}
    \centering
    \includegraphics[width=\linewidth]{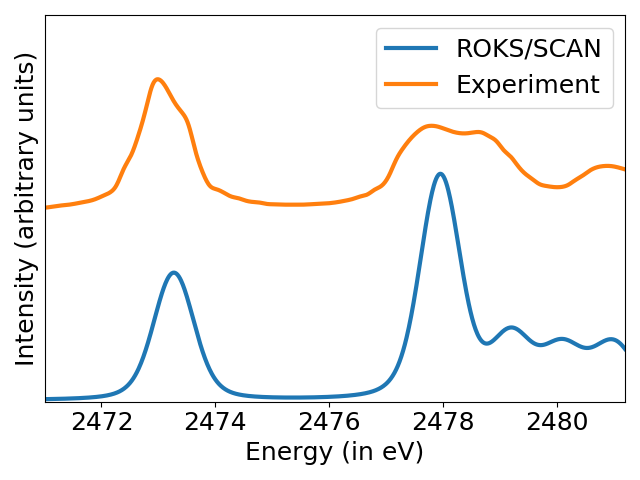}
    \subcaption{S K-edge XAS of gaseous \ce{SO2}\cite{reynaud1996double}, using a doubly augmented (d-aug-) pcX-2 basis on all atoms.}
    \label{fig:so2}
\end{minipage}
\begin{minipage}{0.48\textwidth}
    \centering
    \includegraphics[width=\linewidth]{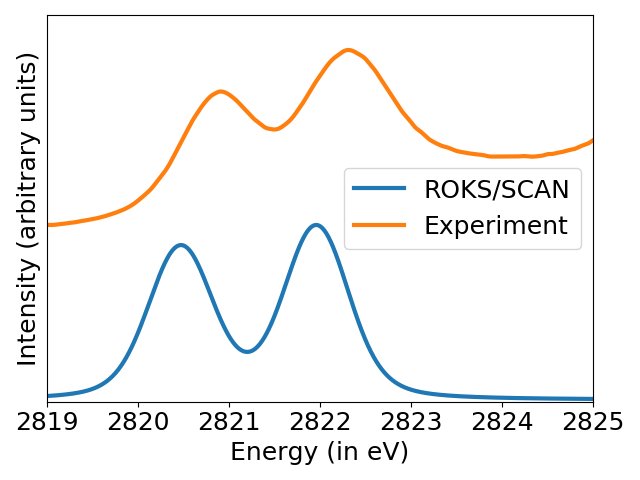}
    \subcaption{Cl K-edge XAS of solid \ce{[Ph4P]2TiCl6}\cite{minasian2012determining}. The system was approximated with a \ce{TiCl6^{ 2-}} ion, placed in a IEF-PCM dielectric utilizing NaCl parameters ($\epsilon_r=6,n=1.5$). A mixed basis (aug-pcX-2 on target Cl, aug-pcseg-1 on other atoms) was used.}
    \label{fig:ticl6}
\end{minipage}\hfill 
\begin{minipage}{0.48\textwidth}
    \centering
    \includegraphics[width=\linewidth]{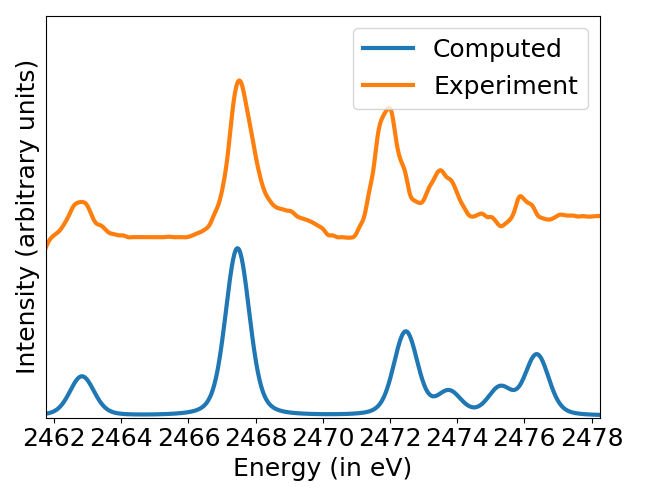}
    \subcaption{Combined S K-edge XAS ($>2470$ eV) and S K$\beta$ emission ($<2470$ eV) of gaseous \ce{CSO}\cite{perera1984molecular}. ROKS was utilized for the XAS and unrestricted $\Delta$SCF for the emission. A doubly augmented (d-aug-) pcX-2 basis on all atoms was used.}
    \label{fig:cso}
\end{minipage}
\caption{Comparison of experimental K-edge spectra for third period elements, and those computed with SCAN. Computed spectra were broadened with a Voigt profile with a Gaussian standard deviation of 0.3 eV and Lorentzian $\gamma$ = 0.121 eV.}
\label{fig:spectrum}
\end{figure}

It is also instructive to look beyond computed excitation energies and consider the full spectrum. Fig \ref{fig:spectrum} presents some examples where the experimental spectrum is compared to computed ones, without using any empirical translations. The peak energies align quite well, within the expected error range of $\sim$ 0.5 eV. The peak heights agree less well, although in some cases this is clearly due to the experimental peaks having different widths (such as in Fig \ref{fig:so2} for \ce{SO2}), vs the uniform broadening utilized for computed spectra. It would be interesting to compute linewidths directly from OO-DFT and determine if that leads to better agreement between theory and experiment. We note that we computed intensities within the dipole-approximation (using \textcolor{black}{transition dipole moments} calculated in the manner described in \textcolor{black}{the supporting information and utilizing non-orthogonal configuration interaction techniques\cite{thom2009hartree}}). It is quite possible that higher order terms have a nonnegligible impact on the experimental X-ray spectrum\cite{list2020beyond}.  Nonetheless, it appears that OO-DFT/X2C is quite effective at reproducing experimental spectra for third period elements. 

\begin{table}[htb!]
\begin{minipage}[b]{0.48\textwidth}
\centering
\begin{tabular}{llrr} \hline\hline
              & \multicolumn{1}{l}{Experiment} & \multicolumn{1}{l}{SCAN} & \multicolumn{1}{l}{SCANh} \\
\ce{TiCl4}         & 4969.2 \cite{debeer2005metal}                      & 4968.9                   & 4969.4                    \\
\ce{TiCpCl3}         &        4968.1 \cite{debeer2005metal}                 &      4968.1              &             4968.5        \\
\ce{TiCp2Cl2}         &             4967.3 \cite{debeer2005metal}           &       4967.5             &            4968.0         \\
\ce{VO(acac)2}     & 5468.4   \cite{rees2016experimental}                      & 5468.6                   & 5469.0     \\
\ce{VCp2Cl2}       & 5468.4     \cite{rees2016experimental}                    & 5468.3                   & 5468.8     \\
\ce{CrO4^{2-}}         &          5996.5   \cite{farges2009chromium}            &        5996.3            &               5996.9    \\
\ce{Cr2O7^{2-}}         &            5996.6 \cite{farges2009chromium}             &   5996.2                 &         5996.7             \\
\ce{MnO4^{-}}         &            6543.3     \cite{hall2014valence}        &          6546.1          &       6546.6              \\
\ce{FeCp2}         & 7111.9      \cite{lancaster2011kbeta}                   & 7116.5                   & 7116.8                    \\
\ce{FeCl4^{-}}     & 7113.2          \cite{westre1997multiplet}              & 7117.1                   & 7117.6                    \\
\textcolor{black}{\ce{CoCl4^{2-}}}  & \textcolor{black}{7709.2    \cite{liu2011speciation}}                     & \textcolor{black}{7714.3}                   & \textcolor{black}{7714.7}\\
\ce{CuCl4^{2-}}  & 8977.6    \cite{dimucci2019myth}                     & 8986.6                   & 8986.8 \\
\ce{Cu(CF3)4^{-}} & 8981.8    \cite{dimucci2019myth}                     & 8990.4                   & 8990.4 \\\hline \hline
\end{tabular}
\subcaption{Individual values (in eV)}
\label{tab:indval}
\end{minipage}\hfill 
\begin{minipage}[b]{0.48\linewidth}
\centering
\includegraphics[width=\columnwidth]{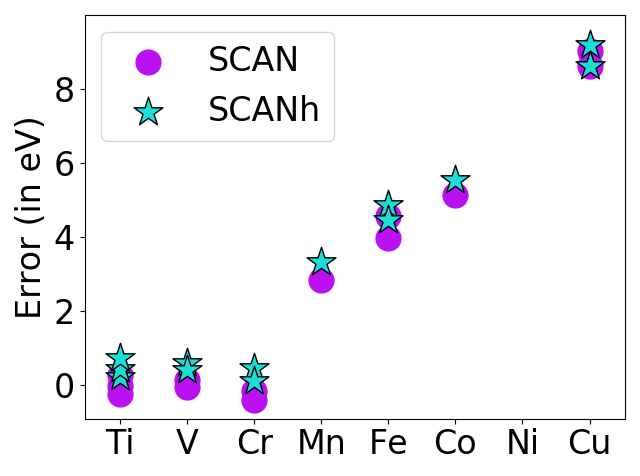}
\subcaption{Excitation energy error vs nuclear charge.}
\label{fig:metal_k}
\end{minipage}
\caption{Lowest symmetry allowed transition metal K-edge transitions. $\Delta$SCF was used for 1s$\to$SOMO transitions of open-shell systems like \ce{VO(acac)2}, while ROKS was used for closed-shell species like \ce{TiCl4}. A mixed basis (decontracted aug-cc-p$\omega$CVTZ\cite{balabanov2005systematically} on excitation site, aug-cc-pVDZ on all other atoms) was utilized.  Experimental data corresponds to the solid state for all species other than TiCl$_4$, whose spectrum was collected in toluene solution.}\label{tab:tmkedge}
\end{table}

We next shift our attention to the 3d transition metals. Transition metal complexes are often open-shell, and have several low-lying orbitals that core electrons can be excited to. K-edge spectra of these species are therefore widely studied, despite the 1s$\to$3d transition in bare atoms being dipole forbidden. Indeed, the transition in centrosymmetric entities (like octahedral complexes) can only be driven by electric quadrupole terms or vibrational symmetry breaking\cite{westre1997multiplet,yamamoto2008assignment}. Tetrahedral complexes however can have some p-d mixing, leading to some dipole allowed intensity for K-edge transitions\cite{yamamoto2008assignment}. Furthermore, X-ray emission spectra (XES) can be collected for $2p/3p\to 1s$ dexcitations, yielding further useful information\cite{pollock2015insights}. 

Table \ref{tab:tmkedge} shows the performance of OO-DFT in predicting transition metal K-edge energies, with the SCAN and SCANh functionals. The results are quite adequate for Ti, V, and Cr but performance is significantly degraded for heavier transition metals like Mn, Fe\textcolor{black}{, Co} and Cu due to significant overestimation (as made evident by Table \ref{fig:metal_k}). This is quite interesting, as SCAN had a slight penchant for underestimation when it came to lighter elements, unlike the significant overbinding observed here for Cu\textcolor{black}{, Co} and Fe (beyond the scale of typical OO-DFT errors). \textcolor{black}{It should be noted that many of the species in Table \ref{tab:tmkedge} are high-spin tetrahedral complexes like \ce{FeCl4^{-}},\ce{CoCl4^{2-}} and \ce{CuCl4^{2-}} or $d^0$ species like \ce{TiCl4}, \ce{CrO4^{2-}} and \ce{MnO4^-}. These species are therefore not particularly multireference, removing one possible source of the error}.

The steep increase in the error with increasing $Z$ appears to suggest a relativistic origin. Indeed, the spin-free one-electron X2C model employed in our work is hardly complete, as it does not include vector (spin-orbit) terms, finite nucleus size effects, or any relativistic contributions to the two-electron terms. The spin-orbit terms are unlikely to be relevant for K-edges, and the finite nucleus size effects are quite small for such elements (being only 0.3 eV for Kr\cite{Southworth2019rel} and therefore smaller for lighter elements). The missing two-electron pieces however are of a comparable magnitude to the error. Specifically, X2C does not transform the electron-electron repulsion terms (the so-called ``picture-change"\cite{saue2011primer}), incorporate additional (Breit)\cite{breit1932dirac} contributions to the electron-electron interaction, or account for quantum electrodynamic effects. These additional terms were previously found to sum to $\sim$ 20 eV of underbinding for the Kr K-edge\cite{Southworth2019rel,niskanen2011relativistic,koziol2018qed}, indicating that overestimation by a few eVs is quite possible for heavier fourth period elements from lack of such effects alone.

We attempted to estimate of the magnitude of missing effects via four-component Dirac-HF (DHF) calculations to account for the picture change and Breit terms, and a QED correction via extrapolation from data in Ref.~\citenum{koziol2018qed}. Specifically, we obtain a Koopmans-level estimate\cite{szabo2012modern} for the first two quantities via computing the difference between the spin-free one-electron X2C 1s eigenvalue and the corresponding DHF eigenvalue for 3d transition metal atomic cations (listed in the supporting information). This analysis reveals that the spin-free X2C one-electron Hamiltonian overbinds by 10.3 eV for \ce{Cu^2+}, which is comparable to the errors seen in Table \ref{tab:tmkedge}. However, these contributions amount to 3.9 eV for \ce{Ti^4+}, indicating that they are very much relevant even for lighter elements (including the third period elements Si--Cl\cite{zheng2022benchmark}). It therefore appears that the good performance of OO-DFT/X2C in this regime is partly due to  cancellation of errors between the missing relativistic contributions and the functional error from SCAN. The sudden increase in error on going from Cr to Mn therefore appears to stem from a breakdown of the error cancellation. The somewhat dramatic increase in error is surprising, but experimental uncertainties in this regime are usually $\sim$ 1 eV and $Z$ is a discrete, integer-valued variable, making such sudden increases possible. 
It should also be noted that the missing relativistic corrections discussed here are estimated at a Koopmans' theorem based level and so can differ somewhat from a full OO-DFT estimate (which we cannot find at present).

However, it is worth stressing that the X2C model nonetheless manages to account for the vast majority of relativistic effects for even species like Cu/Kr. It is also worth noting that translating OO-DFT/X2C metal K-edge spectra for alignment with experiment would involve much smaller shifts than TDDFT, reducing the magnitude of potential translation driven error. Nonetheless, we wish to avoid any need for empirical translation of spectra, and instead intend to pursue more accurate relativistic models to quantitatively model the K-edge spectra of heavier elements.

\begin{table}[htb!]
\begin{tabular}{lrrrrrr} \hline\hline
            & \multicolumn{3}{l}{Experiment}                                                       & \multicolumn{3}{l}{Theoretical L$_3$ edge}                                             \\
            & \multicolumn{1}{l}{L$_3$} & \multicolumn{1}{l}{L$_2$} & \multicolumn{1}{l}{3J$=$L$_2$-L$_3$} & \multicolumn{1}{l}{SCAN} & \multicolumn{1}{l}{SCANh} & \multicolumn{1}{l}{SCAN0} \\ \hline
\ce{TiCl4}       & 456.9\cite{wen1993inner}                 & 462.5                  & 5.6                                & 455.9                    & 456.4                     & 457.1                     \\
\ce{Mn(OH2)6^{2+}}    & 639.7\cite{mitzner2013edge}                  & 649.1                  & 9.4                                & 638.7                    & 638.9                     & 639.3                     \\
\ce{Fe(CN)6^{3-}}  & 705.8\cite{hocking2006fe}                 & 718.4                  & 12.6                               & 706.1                    & 706.5                     & 707.1                     \\
\ce{FeCp2}       & 708.9\cite{wen1992inner}                  & 721.2                  & 12.3                               & 708.2                    & 708.5                     & 709.0                     \\
\ce{CuCl4$^{2-}$} (D$_{2h}$) & 930.1\cite{dimucci2019myth}                   & 950.1                  & 20                                 & 929.4                    & 929.5                     & 929.8                     \\
\ce{Cu(CF3)4^-}    & 934.7\cite{dimucci2019myth}                  & 954.7                  & 20                                 & 933.5                    & 933.5                     & 933.7                     \\
\hline\hline              
\end{tabular}
\caption{Lowest symmetry allowed L-edge (2p) transitions for transition metal containing species (in eV). $\Delta$SCF was used for 2p$\to$SOMO transitions of open-shell systems like CuCl$_4^{2-}$, while ROKS was used for closed-shell species like ferrocene (FeCp$_2$). Computed multiplet averaged energies (found from averaging over all the 2p orbitals) were red-shifted by $J$ (where $3J$ is the experimental energy gap between the L$_2$ and L$_3$ peaks) to better approximate the experimental L$_3$ peaks. A mixed basis (decontracted aug-cc-p$\omega$CVTZ on excitation site, aug-cc-pVDZ on all other atoms) was utilized for the calculations.}
\label{tab:tmledge}
\end{table}

Metal L-edge (2p) spectra are also extensively studied via experiment. Accurate computation of spectra necessitates going beyond the scalar relativistic paradigm as the degeneracy of the $2p$ levels is broken by spin-orbit coupling. Scalar relativistic models like X2C can potentially yield a reasonable estimate for the multiplet averaged peak, which can be subsequently shifted by atomic spin-orbit values to obtain the experimentally observed $L_{3/2}$ (L$_3$-edge) and $L_{1/2}$ (L$_2$-edge) multiplet peaks. This strategy has been found to be effective for L-edges of Si,P,S and Cl\cite{hait2020highly}. However, it is less appealing for heavier elements due to the larger magnitude of the multiplet splitting (20 eV for Cu vs 1.6 eV for Cl), which can potentially exert a direct influence in the OO process. We have nonetheless applied this measure to compute L-edges for a few 3d metal containing species (shown in Table \ref{tab:tmledge}), to assess this approach. It is quite apparent that the errors are much larger here (relative to Fig \ref{fig:boxplots}), with significant underestimation being the norm for all species aside for [Fe(CN)$_6$]$^{3+}$. The results are also often quite sensitive to \% HF exchange (more so than most of the species considered till this point). Nonetheless, the worst case errors are just slightly above $1$ eV, indicating that OO-DFT with proper inclusion of spin-orbit effects and careful functional choice has the potential to be quite accurate in predicting metal L-edges. We are presently investigating this aspect further.  

Overall, it is clear that the OO-DFT/X2C combination is capable of consistently delivering $<1$ eV error for core-level excitation/ionization energies for third period elements and reproduce experimental spectra fairly well. Indeed, the typical error is $\sim$ 0.5 eV for SCAN/SCANh, as shown by Fig \ref{fig:boxplots} and the RMSEs reported in Tables \ref{tab:gasxps} and \ref{tab:gasxas}.  X2C therefore extends the applicability of OO-DFT methods to elements heavier than F, which was previously the limit for computational core-level spectroscopy with such methods (without post-facto application of ad-hoc relativistic corrections). OO-DFT X2C also appears to be adequate for the K-edges of the transition metals Ti, V and Cr. However, performance degrades starting with Mn, \textcolor{black}{possibly} due to lack of relativistic effects in the two-electron interaction terms. The spin-free X2C model is also incapable of accounting for the spin-orbit splitting observed in experimental L-edge spectra. Accounting for these effects would be critical for extension of OO-DFT beyond the cases explored in this work. We are also attempting to apply OO-DFT extensively to K-edge spectra of third period elements, in order to uncover any additional limitations of the approach. Any resulting insight could also prove valuable in training density functionals that are accurate in modeling both ground states and OO-DFT excited states, as it is quite possible that not much further improvement in prediction quality can be obtained from functionals trained solely for the ground state. 

\section*{Computational methods}
All calculations were performed with a development version of the Q-Chem 5.4 package\cite{epifanovsky2021software} and these new capabilities will be publicly available with the next release of code. Local exchange–correlation integrals for DFT were calculated over a radial grid with 99 points and an angular Lebedev grid with 590 points. Ref \citenum{hait2020accurate} lays out the protocol for computing $\Delta$SCF excited states, while Ref \citenum{hait2020highly} does the same for ROKS. The restricted open-shell optimizations necessary in this process were performed through square gradient minimization (SGM\cite{hait2020excited}) and unrestricted optimizations with initial maximum overlap method (IMOM\cite{barca2018simple}). The core-hole was localized onto a single atom for species with equivalent atoms (like S in \ce{CS2}), in order to prevent errors arising from delocalization\cite{perdew1982density,hait2018delocalization} of the hole over multiple sites\cite{hait2020highly}. Standard values of dielectric constant $\epsilon_r$ and refractive index $n$ were used for IEF-PCM modeling of common solvents like cyclohexane (listed in the supporting information). The corresponding data for most solid state materials was not available, and consequently all solid state environments
were modeled with NaCl parameters ($\epsilon_r=6,n=1.5$). This should be reasonable for ionic solids, although perhaps a little too polar for molecular solids like \ce{OPPh3} (Table \ref{tab:largegas}). However, the resulting values for  \ce{OPPh3} were quite close to vacuum calculations (as shown in the supporting information), suggesting that the environment exerted negligible impact on the spectrum. Four-component DHF calculations for 3d transition metal ions were carried out within the PySCF package.\cite{sun2020recent}

Experimental geometries were used whenever possible, through gas phase data from NIST\cite{johnson2015nist} or crystal structures from the Cambridge structural database\cite{groom2016cambridge}. Structures were optimized with $\omega$B97M-V\cite{wB97MV}/aug-pcseg-1 under gas phase conditions, if experimental data was unavailable.  
The source of all of the geometries is listed in the supporting information, along with the associated atomic coordinates. In particular, the \ce{CuCl4}$^{2-}$ ion studied is of $D_{2h}$ (distorted tetrahedral) symmetry, corresponding to \ce{Cs2CuCl4}\cite{mcginnety1972cesium}. \textcolor{black}{The ground state geometries were employed for excited state calculations, consistent with the Franck-Condon principle\cite{franck1926elementary,condon1926theory}.}
\section*{Acknowledgment} 
This work was supported by the Liquid Sunlight Alliance, which is funded by the U.S. Department of Energy, Office of Science, Office of Basic Energy Sciences, Fuels from Sunlight Hub under Award Number DE-SC0021266 with additional support from the Director, Office of Science, Office of Basic Energy Sciences, of the U.S. Department of Energy, under Contract No. DE-AC02-05CH11231. We would like to thank Prof. Lan Cheng for helpful discussions. 

\section*{Supporting Information}
PDF: X2C transformation, validation, short note about S K-edge binding energies of some species, short note on computing transition dipole moments. \\
XLXS: Basis set convergence, light element data, functional screening information, X2C validation, post spin-free X2C one electron relativistic contributions.\\
ZIP: Geometries of all species considered in xyz format (along with provenance).
\section*{Conflicts of Interest}
M.H.-G. is a part-owner of Q-Chem, which is the software platform in which the developments described here were implemented.

\section{\LARGE Supporting Information}
\section{Exact Two-Component (X2C) Relativistic Hamiltonian}
The starting point for deriving the X2C relativistic model is the four-component one-electron Dirac Hamiltonian (Eq~\ref{eq:dirac}) represented in a restricted kinetic balance (RKB) form\cite{kutzelnigg1984basis}. Our goal is to use a unitary transformation that effectively decouples the positive and negative-energy solutions of the Dirac equation, since we are only interested in describing electrons. In Eq.~\ref{eq:dirac}, $T$, $V$ and $S$ are the usual non-relativistic kinetic energy, nuclear attraction and overlap matrices, respectively, represented in a basis of atomic orbitals $\{ \phi_\mu \} $. $W$ is defined as the matrix representation of the operator in Eq.~\ref{eq:Wx2c}, where $\vec{\sigma}$ is the vector of Pauli matrices, and $\vec{p}$ and $V$ are the momentum and nuclear-attraction operators, respectively. As indicated in Eq~\ref{eq:Wx2c}, $\hat{W}$ can be decomposed into spin-free ($\hat{W}_{\textrm{SF}}$) and spin-orbit ($\hat{W}_{\textrm{SO}}$) components (likewise, the matrix representation $W$ can also be separated into $W_{\textrm{SF}}$ and $W_{\textrm{SO}}$). Finally, the solutions of Eq.~\ref{eq:dirac} are characterized by their large ($C_L$) and small ($C_S$) components.  

\begin{align}
    \begin{bmatrix}
        V & T \\
        T & \dfrac{W}{4c^2} -T \\
    \end{bmatrix}
    \begin{bmatrix}
        C_L \\
        C_S \\
    \end{bmatrix} &= E
        \begin{bmatrix}
        S & 0 \\
        0 & \dfrac{1}{2c^2}T \\
    \end{bmatrix}
    \begin{bmatrix}
        C_L \\
        C_S \\
    \end{bmatrix}\label{eq:dirac}
\\
    \hat{W} &= (\vec{\sigma}\cdot \vec{p}) V  (\vec{\sigma}\cdot\vec{p}) = \left( \vec{p}\cdot V\vec{p} + i \vec{\sigma}\cdot(\vec{p} \times V\vec{p}) \right)=  \hat{W}_\textrm{SF} + i \vec{\sigma} \cdot \hat{W}_\textrm{SO}
\label{eq:Wx2c}\\
    W_{\mu\nu}&=\bra{\phi_\mu}\hat{W}\ket{\phi_\nu}
\end{align}

As outlined in the main text, we focused on the inclusion of scalar relativistic effects to OO-DFT, so only the spin-free $W_\textrm{SF}$ was taken into account and the spin-orbit $W_\textrm{SO}$ ignored. It is possible to evaluate the matrix elements of $W_\textrm{SF}$ analytically in a gaussian type orbital (GTO) basis, but we chose to take a finite difference approach instead, which can be readily utilized by any quantum chemistry code without too much difficulty. 

Our approach is based on the momentum operator being a generator of translation. We know that given a basis function $\phi_\mu(\vec{r})$ (GTO or otherwise), we have:

\begin{align}
    \phi_\mu(\vec{r}+\delta\hat{n})=T(\delta\hat{n})\phi_\mu(\vec{r})
\end{align}
where the translation operator $T(\delta\hat{n})=\exp\left(-i \delta \hat{n}\cdot\vec{p}\right)$ translates the orbital by $\delta$ along the unit vector $\hat{n}$. With this, we can see that:
\begin{align}\label{eq:fd_basis}
    \left(\hat{n}\cdot\vec{p} \right)\phi_\mu(\vec{r}) &=-i\dfrac{\phi_\mu(\vec{r}+\delta\hat{n})-\phi_\mu(\vec{r}-\delta\hat{n})}{2\delta}+O(\delta^2)
\end{align}
which can also be obtained from noting that the momentum operator $\left(\hat{n}\cdot\vec{p} \right)$ along $\hat{n}$ is equivalent to taking the derivative with respect to the spatial coordinate along $\hat{n}$.
We can thus define a finite difference momentum operator $\vec{p}_2(\delta)$ such that:
\begin{align}
    \left(\hat{x}\cdot\vec{p}_2(\delta) \right)\phi_\mu(\vec{r}) &=-i\dfrac{\phi_\mu(\vec{r}+\delta\hat{x})-\phi_\mu(\vec{r}-\delta\hat{x})}{2\delta}\\
    \left(\hat{y}\cdot\vec{p}_2(\delta) \right)\phi_\mu(\vec{r}) &=-i\dfrac{\phi_\mu(\vec{r}+\delta\hat{y})-\phi_\mu(\vec{r}-\delta\hat{y})}{2\delta}\\
    \left(\hat{z}\cdot\vec{p}_2(\delta) \right)\phi_\mu(\vec{r}) &=-i\dfrac{\phi_\mu(\vec{r}+\delta\hat{z})-\phi_\mu(\vec{r}-\delta\hat{z})}{2\delta}
\end{align}
Consequently, an approximate $\hat{W}_\textrm{2,SF}$ can be defined as:
\begin{align}
    \hat{W}_\textrm{2,SF}(\delta)&=\vec{p}_2(\delta)\cdot V\vec{p}_2(\delta)\\
   &= \left(\hat{x}\cdot\vec{p}_2(\delta) \right)V\left(\hat{x}\cdot\vec{p}_2(\delta) \right)+\left(\hat{y}\cdot\vec{p}_2(\delta) \right)V\left(\hat{y}\cdot\vec{p}_2(\delta) \right)+\left(\hat{z}\cdot\vec{p}_2(\delta) \right)V\left(\hat{z}\cdot\vec{p}_2(\delta) \right)
\end{align}
The resulting matrix elements can be found in the following manner:
\begin{align}
   &\phi^*_\mu(\vec{r}) \left(\hat{x}\cdot\vec{p}_2(\delta) \right)V\left(\hat{x}\cdot\vec{p}_2(\delta)\right) \phi_\nu(\vec{r})\notag\\=&
   \left(\hat{x}\cdot\vec{p}_2(\delta)\phi_\mu(\vec{r}) \right)^*V\left(\hat{x}\cdot\vec{p}_2(\delta)\phi_\nu(\vec{r}) \right)\\
   =\notag&\dfrac{1}{4\delta^2}\left(\phi_\mu(\vec{r}+\delta\hat{x})V\phi_\nu(\vec{r}+\delta\hat{x})+\phi_\mu(\vec{r}-\delta\hat{x})V\phi_\nu(\vec{r}-\delta\hat{x})\right. \\&\left.-\phi_\mu(\vec{r}-\delta\hat{x})V\phi_\nu(\vec{r}+\delta\hat{x})-\phi_\mu(\vec{r}+\delta\hat{x})V\phi_\nu(\vec{r}-\delta\hat{x})\right)
\end{align}
Thus matrix elements of $W_\textrm{2,SF}$ can be found via evaluation of matrix elements of $V$ using slightly translated GTOs $\phi_\mu(\vec{r}+\delta\hat{x})$ etc, which should be not too challenging for any quantum chemistry code. The resulting values will however have some finite difference error as:
\begin{align}
    W_\textrm{SF}&=W_\textrm{2,SF}(\delta)+O(\delta^2)
\end{align}
This error can be greatly reduced by utilizing different values of $\delta$ and refining $W_\text{SF}$ via:
\begin{align}
     W_\textrm{SF}&=\dfrac{W_\textrm{2,SF}(2\delta)-16W_\textrm{2,SF}(\delta)+64W_\textrm{2,SF}\left(\dfrac{\delta}{2}\right)}{45}+O(\delta^6) \label{eq:stencil}
\end{align}
or other higher order stencils. We used Eq \ref{eq:stencil} in our implementation with $\delta=1\times 10^{-4}$ atomic units, 
resulting in negligible finite difference error.

One can obtain the coupling matrix $X$ for the positive energy solutions of Eq.~\ref{eq:dirac} as the ratio between the large and small components, as indicated in Eq.~\ref{eq:coupling_x}. The renormalization matrix $R$ is then defined as in Eq.~\ref{eq:renorm_r}, where $\tilde{S}$ (Eq.~\ref{eq:stilde}) is a modified overlap matrix that takes into account the folding of the small component into the large one.

\begin{align}\label{eq:coupling_x}
    X &= C_S (C_L)^{-1}\\
\label{eq:renorm_r}
    R &= S^{-1/2}\left( S^{-1/2} \tilde{S} S^{-1/2}\right)^{-1/2}S^{1/2}\\
\label{eq:stilde}
    \tilde{S} &= S + X^\dagger \frac{1}{2c^2}X
\end{align}

With the $X$ and $R$ matrices, we can now calculate the effective SF-X2C1e kinetic energy (Eq.~\ref{eq:h_x2c}) and nuclear attraction (Eq.~\ref{eq:v_x2c}) operators for subsequent KS-DFT calculations.

\begin{align}
\label{eq:h_x2c}
    T_{X2C} = R^\dagger\left(TX + X^\dagger T - X^\dagger T X\right) R\\
\label{eq:v_x2c}
    V_{X2C} = R^\dagger\left(V + \frac{1}{4c^2}X^\dagger W_{SF} X\right) R
\end{align}

We checked our finite differences implementation with the analytical one from the Psi4 package \cite{smith2020psi4} by comparing absolute Hartree-Fock (HF) energies. Table~\ref{tab:hf_psi4} indicates that the energy difference between the two implementations is usually on the order of $10^{-7}$ hartrees or less (which should have negligible impact on core-level spectrum calculations).

\begin{table}[!htp]
\footnotesize
\begin{tabular}{lllllll}
\textbf{}       & \multicolumn{3}{c}{\textbf{Non-relativistic   ground state total energy}} & \multicolumn{3}{c}{\textbf{X2C ground state total   energy}} \\
\textbf{System} & \textbf{Psi4}          & \textbf{Q-Chem}       & \textbf{Difference}      & \textbf{Psi4}     & \textbf{Q-Chem}   & \textbf{Difference}  \\
\ce{SiH4}       & -291.2641256030        & -291.2641256030       & -2.0E-11                 & -291.8673077884   & -291.8673077905   & -2.1E-09             \\
\ce{PH3}        & -342.4903826664        & -342.4903826663       & 7.0E-11                  & -343.3071475141   & -343.3071475130   & 1.1E-09              \\
\ce{H2S}        & -398.7160009253        & -398.7160009254       & -7.0E-11                 & -399.7996418715   & -399.7996419031   & -3.2E-08             \\
\ce{HCl}        & -460.1087555150        & -460.1087555151       & -1.5E-10                 & -461.5217242958   & -461.5217242810   & 1.5E-08              \\
\ce{Ar}         & -526.8134620705        & -526.8134620705       & 2.0E-11                  & -528.6278041021   & -528.6278041049   & -2.8E-09             \\
\ce{TiCl4}      & -2686.7122452582       & -2686.7122452457      & 1.3E-08                  & -2696.7140080590  & -2696.7140080397  & 1.9E-08              \\
\ce{CrO4^2-}    & -1342.4736099183       & -1342.4736099171      & 1.2E-09                  & -1349.0259066545  & -1349.0259066508  & 3.7E-09              \\
\ce{MnO4^-}     & -1448.7946004947       & -1448.7946004920      & 2.7E-09                  & -1456.5639058910  & -1456.5639058892  & 1.8E-09              \\
\ce{FeCl4^2-}   & -3100.7604247005       & -3100.7604246828      & 1.8E-08                  & -3115.3237860409  & -3115.3237860275  & 1.3E-08              \\
\ce{CoCl4^2-}   & -3219.7195689779       & -3219.7195689046      & 7.3E-08                  & -3235.8506032331  & -3235.8506031501  & 8.3E-08              \\
\ce{NiCl4^2-}   & -3345.1648205253       & -3345.1648204645      & 6.1E-08                  & -3363.0642419690  & -3363.0642419050  & 6.4E-08              \\
\ce{CuCl4^2-}   & -3477.2188051504       & -3477.2188050680      & 8.2E-08                  & -3497.1060177623  & -3497.1060176503  & 1.1E-07             
\end{tabular}
\caption{Absolute ground state HF energies obtained with Psi4's analytical implementation of X2C and Q-Chem's finite differences. A comparison of non-relativistic energies is also supplied for comparison. The decontracted aug-cc-pCVTZ basis was used for \ce{SiH4},\ce{PH3},\ce{HCl} and \ce{Ar}; while decontracted cc-pVDZ was used for the transition metal containing systems.}
\label{tab:hf_psi4}
\end{table}

\newpage 
\section{Note on binding energies for some S compounds}
The 1s binding energies for several S compounds in Table 1 of the main paper were obtained from Ref \citenum{sodhi1986kll}, which reported how much those energies differed from the 1s ionization energy of \ce{H2S}. However, the reference energy for \ce{H2S} reported by Ref \citenum{sodhi1986kll} was in error by 0.5 eV, as a relativistic effect was missed during calibration of experiment\cite{carroll1987relativistic,cavell1987effect}. We therefore used the \ce{H2S} 1s binding energy from Ref \citenum{keski1976energies} (which was validated by subsequent studies\cite{carroll1987relativistic,cavell1987effect}) as the reference, and added the shifts reported by Ref \citenum{sodhi1986kll} to obtain the true binding energies. This protocol should be acceptable, as the shifts reported in  Ref \citenum{sodhi1986kll} should not be affected by the error in the reference value\cite{carroll1987relativistic,cavell1987effect}. 

\newpage 
\section{Note on computing transition dipole moments}
The transition dipole moments for excitations described in this work were computed by treating the KS Slater determinants as pseudo-wavefunctions. In other words, if the ground state of a species has a KS determinant $\ket{\Phi_0}$ and the $\Delta$SCF excited state has a KS determinant $\ket{\Phi_n}$, the transition dipole moment between them is given by:
\begin{align}
    \vec{\mu}_{0n}&=\bra{\Phi_0}\hat{\vec{\mu}}\ket{\Phi_n}
\end{align}
where $\hat{\vec{\mu}}$ is the dipole operator. 

For ROKS, there are two excited state determinants with equal weight (corresponding to the exchange of up and down spins). If those determinants are $\ket{\Phi_n}$ and $\ket{\bar{\Phi}_n}$ respectively, the ROKS `wave function' is $\dfrac{\ket{\Phi_n}+\ket{\bar{\Phi}_n}}{\sqrt{2}}$, leading to a transition dipole of:
\begin{align}
    \vec{\mu}_{0n}&=\dfrac{1}{\sqrt{2}}\left(\bra{\Phi_0}\hat{\vec{\mu}}\ket{\Phi_n}+\bra{\Phi_0}\hat{\vec{\mu}}\ket{\bar{\Phi}_n}\right)\\
    &=\sqrt{2}\bra{\Phi_0}\hat{\vec{\mu}}\ket{\Phi_n}
\end{align}
if the ground state $\ket{\Phi_0}$ is closed-shell (from spin-inversion symmetry, as  $\hat{\vec{\mu}}$ is purely a spatial one-particle operator). 

Computation of $\bra{\Phi_0}\hat{\vec{\mu}}\ket{\Phi_n}$ is not quite straightforward as the two determinants are constructed from different (non-orthogonal) sets of orbitals. Non-orthogonal configuration interaction techniques (as described in Ref  \citenum{thom2009hartree}) have to be therefore employed, incurring some extra computational cost vs cases with orthogonal orbitals, but ultimately having negligible effect relative to the cost of excited state orbital optimization. 

One further point to note is that $\vec{\mu}_{0n}$ is not translationally invariant for cases with $\braket{\Phi_0}{\Phi_n}\ne 0$. This is easily resolved in neutral systems via inclusion of the nuclear contribution to the $\hat{\vec{\mu}}$ operator, which restores translational invariance (and which amounts to translating the system to the center of nuclear charge). For charged systems, it is necessary to directly translate the system to the center of nuclear charge in order to have results that are not heavily contaminated by spurious contributions from the nonzero overlap $\braket{\Phi_0}{\Phi_n}$. Alternatively, symmetric orthogonalization \cite{bourne2021reliable} can be carried out to obtain translationally invariant results. Both routes give similar results for valence $\Delta$SCF on neutral systems\cite{bourne2021reliable} and we have therefore used the simpler route of operating from the center of nuclear charge for the small systems examined in Fig 4 of the main manuscript.

\bibliography{references}

\providecommand{\latin}[1]{#1}
\makeatletter
\providecommand{\doi}
  {\begingroup\let\do\@makeother\dospecials
  \catcode`\{=1 \catcode`\}=2 \doi@aux}
\providecommand{\doi@aux}[1]{\endgroup\texttt{#1}}
\makeatother
\providecommand*\mcitethebibliography{\thebibliography}
\csname @ifundefined\endcsname{endmcitethebibliography}
  {\let\endmcitethebibliography\endthebibliography}{}
\begin{mcitethebibliography}{142}
\providecommand*\natexlab[1]{#1}
\providecommand*\mciteSetBstSublistMode[1]{}
\providecommand*\mciteSetBstMaxWidthForm[2]{}
\providecommand*\mciteBstWouldAddEndPuncttrue
  {\def\EndOfBibitem{\unskip.}}
\providecommand*\mciteBstWouldAddEndPunctfalse
  {\let\EndOfBibitem\relax}
\providecommand*\mciteSetBstMidEndSepPunct[3]{}
\providecommand*\mciteSetBstSublistLabelBeginEnd[3]{}
\providecommand*\EndOfBibitem{}
\mciteSetBstSublistMode{f}
\mciteSetBstMaxWidthForm{subitem}{(\alph{mcitesubitemcount})}
\mciteSetBstSublistLabelBeginEnd
  {\mcitemaxwidthsubitemform\space}
  {\relax}
  {\relax}

\bibitem[Yuhas \latin{et~al.}(2007)Yuhas, Fakra, Marcus, and
  Yang]{yuhas2007probing}
Yuhas,~B.~D.; Fakra,~S.; Marcus,~M.~A.; Yang,~P. Probing the local coordination
  environment for transition metal dopants in zinc oxide nanowires. \emph{Nano
  Lett.} \textbf{2007}, \emph{7}, 905--909\relax
\mciteBstWouldAddEndPuncttrue
\mciteSetBstMidEndSepPunct{\mcitedefaultmidpunct}
{\mcitedefaultendpunct}{\mcitedefaultseppunct}\relax
\EndOfBibitem
\bibitem[Pollock and DeBeer(2015)Pollock, and DeBeer]{pollock2015insights}
Pollock,~C.~J.; DeBeer,~S. Insights into the geometric and electronic structure
  of transition metal centers from valence-to-core X-ray emission spectroscopy.
  \emph{Acc. Chem. Res.} \textbf{2015}, \emph{48}, 2967--2975\relax
\mciteBstWouldAddEndPuncttrue
\mciteSetBstMidEndSepPunct{\mcitedefaultmidpunct}
{\mcitedefaultendpunct}{\mcitedefaultseppunct}\relax
\EndOfBibitem
\bibitem[Westre \latin{et~al.}(1997)Westre, Kennepohl, DeWitt, Hedman, Hodgson,
  and Solomon]{westre1997multiplet}
Westre,~T.~E.; Kennepohl,~P.; DeWitt,~J.~G.; Hedman,~B.; Hodgson,~K.~O.;
  Solomon,~E.~I. A multiplet analysis of Fe K-edge 1s→ 3d pre-edge features
  of iron complexes. \emph{J. Am. Chem. Soc.} \textbf{1997}, \emph{119},
  6297--6314\relax
\mciteBstWouldAddEndPuncttrue
\mciteSetBstMidEndSepPunct{\mcitedefaultmidpunct}
{\mcitedefaultendpunct}{\mcitedefaultseppunct}\relax
\EndOfBibitem
\bibitem[Solomon \latin{et~al.}(2005)Solomon, Hedman, Hodgson, Dey, and
  Szilagyi]{solomon2005ligand}
Solomon,~E.~I.; Hedman,~B.; Hodgson,~K.~O.; Dey,~A.; Szilagyi,~R.~K. Ligand
  K-edge X-ray absorption spectroscopy: covalency of ligand--metal bonds.
  \emph{Coord. Chem. Rev.} \textbf{2005}, \emph{249}, 97--129\relax
\mciteBstWouldAddEndPuncttrue
\mciteSetBstMidEndSepPunct{\mcitedefaultmidpunct}
{\mcitedefaultendpunct}{\mcitedefaultseppunct}\relax
\EndOfBibitem
\bibitem[Kubin \latin{et~al.}(2018)Kubin, Guo, Kroll, L{\"o}chel, K{\"a}llman,
  Baker, Mitzner, Gul, Kern, F{\"o}hlisch, \latin{et~al.}
  others]{kubin2018probing}
Kubin,~M.; Guo,~M.; Kroll,~T.; L{\"o}chel,~H.; K{\"a}llman,~E.; Baker,~M.~L.;
  Mitzner,~R.; Gul,~S.; Kern,~J.; F{\"o}hlisch,~A. \latin{et~al.}  Probing the
  oxidation state of transition metal complexes: a case study on how charge and
  spin densities determine Mn L-edge X-ray absorption energies. \emph{Chem.
  Sci.} \textbf{2018}, \emph{9}, 6813--6829\relax
\mciteBstWouldAddEndPuncttrue
\mciteSetBstMidEndSepPunct{\mcitedefaultmidpunct}
{\mcitedefaultendpunct}{\mcitedefaultseppunct}\relax
\EndOfBibitem
\bibitem[Chergui and Collet(2017)Chergui, and Collet]{chergui2017photoinduced}
Chergui,~M.; Collet,~E. Photoinduced structural dynamics of molecular systems
  mapped by time-resolved X-ray methods. \emph{Chem. Rev.} \textbf{2017},
  \emph{117}, 11025--11065\relax
\mciteBstWouldAddEndPuncttrue
\mciteSetBstMidEndSepPunct{\mcitedefaultmidpunct}
{\mcitedefaultendpunct}{\mcitedefaultseppunct}\relax
\EndOfBibitem
\bibitem[Bhattacherjee and Leone(2018)Bhattacherjee, and
  Leone]{bhattacherjee2018ultrafast}
Bhattacherjee,~A.; Leone,~S.~R. Ultrafast X-ray Transient Absorption
  Spectroscopy of Gas-Phase Photochemical Reactions: A New Universal Probe of
  Photoinduced Molecular Dynamics. \emph{Acc. Chem. Res.} \textbf{2018},
  \emph{51}, 3203--3211\relax
\mciteBstWouldAddEndPuncttrue
\mciteSetBstMidEndSepPunct{\mcitedefaultmidpunct}
{\mcitedefaultendpunct}{\mcitedefaultseppunct}\relax
\EndOfBibitem
\bibitem[Kraus \latin{et~al.}(2018)Kraus, Z{\"u}rch, Cushing, Neumark, and
  Leone]{kraus2018ultrafast}
Kraus,~P.~M.; Z{\"u}rch,~M.; Cushing,~S.~K.; Neumark,~D.~M.; Leone,~S.~R. The
  ultrafast X-ray spectroscopic revolution in chemical dynamics. \emph{Nat.
  Rev. Chem.} \textbf{2018}, \emph{2}, 82--94\relax
\mciteBstWouldAddEndPuncttrue
\mciteSetBstMidEndSepPunct{\mcitedefaultmidpunct}
{\mcitedefaultendpunct}{\mcitedefaultseppunct}\relax
\EndOfBibitem
\bibitem[Ochmann \latin{et~al.}(2017)Ochmann, Von~Ahnen, Cordones, Hussain,
  Lee, Hong, Adamczyk, Vendrell, Kim, Schoenlein, \latin{et~al.}
  others]{ochmann2017light}
Ochmann,~M.; Von~Ahnen,~I.; Cordones,~A.~A.; Hussain,~A.; Lee,~J.~H.; Hong,~K.;
  Adamczyk,~K.; Vendrell,~O.; Kim,~T.~K.; Schoenlein,~R.~W. \latin{et~al.}
  Light-induced radical formation and isomerization of an aromatic thiol in
  solution followed by time-resolved x-ray absorption spectroscopy at the
  sulfur K-edge. \emph{J. Am. Chem. Soc.} \textbf{2017}, \emph{139},
  4797--4804\relax
\mciteBstWouldAddEndPuncttrue
\mciteSetBstMidEndSepPunct{\mcitedefaultmidpunct}
{\mcitedefaultendpunct}{\mcitedefaultseppunct}\relax
\EndOfBibitem
\bibitem[Bhattacherjee \latin{et~al.}(2017)Bhattacherjee, Pemmaraju, Schnorr,
  Attar, and Leone]{bhattacherjee2017ultrafast}
Bhattacherjee,~A.; Pemmaraju,~C.~D.; Schnorr,~K.; Attar,~A.~R.; Leone,~S.~R.
  Ultrafast intersystem crossing in acetylacetone via femtosecond x-ray
  transient absorption at the carbon K-edge. \emph{J. Am. Chem. Soc.}
  \textbf{2017}, \emph{139}, 16576--16583\relax
\mciteBstWouldAddEndPuncttrue
\mciteSetBstMidEndSepPunct{\mcitedefaultmidpunct}
{\mcitedefaultendpunct}{\mcitedefaultseppunct}\relax
\EndOfBibitem
\bibitem[Dreuw and Head-Gordon(2005)Dreuw, and Head-Gordon]{dreuw2005single}
Dreuw,~A.; Head-Gordon,~M. {Single-reference ab initio methods for the
  calculation of excited states of large molecules}. \emph{Chem. Rev.}
  \textbf{2005}, \emph{105}, 4009--4037\relax
\mciteBstWouldAddEndPuncttrue
\mciteSetBstMidEndSepPunct{\mcitedefaultmidpunct}
{\mcitedefaultendpunct}{\mcitedefaultseppunct}\relax
\EndOfBibitem
\bibitem[Krylov(2008)]{krylov2008equation}
Krylov,~A.~I. Equation-of-motion coupled-cluster methods for open-shell and
  electronically excited species: The hitchhiker's guide to Fock space.
  \emph{Annu. Rev. Phys. Chem.} \textbf{2008}, \emph{59}, 433--462\relax
\mciteBstWouldAddEndPuncttrue
\mciteSetBstMidEndSepPunct{\mcitedefaultmidpunct}
{\mcitedefaultendpunct}{\mcitedefaultseppunct}\relax
\EndOfBibitem
\bibitem[Runge and Gross(1984)Runge, and Gross]{runge1984density}
Runge,~E.; Gross,~E. K.~U. Density-functional theory for time-dependent
  systems. \emph{Phys. Rev. Lett.} \textbf{1984}, \emph{52}, 997--1000\relax
\mciteBstWouldAddEndPuncttrue
\mciteSetBstMidEndSepPunct{\mcitedefaultmidpunct}
{\mcitedefaultendpunct}{\mcitedefaultseppunct}\relax
\EndOfBibitem
\bibitem[Wenzel \latin{et~al.}(2014)Wenzel, Wormit, and
  Dreuw]{wenzel2014calculating}
Wenzel,~J.; Wormit,~M.; Dreuw,~A. Calculating core-level excitations and x-ray
  absorption spectra of medium-sized closed-shell molecules with the
  algebraic-diagrammatic construction scheme for the polarization propagator.
  \emph{J. Comput. Chem.} \textbf{2014}, \emph{35}, 1900--1915\relax
\mciteBstWouldAddEndPuncttrue
\mciteSetBstMidEndSepPunct{\mcitedefaultmidpunct}
{\mcitedefaultendpunct}{\mcitedefaultseppunct}\relax
\EndOfBibitem
\bibitem[Lopata \latin{et~al.}(2012)Lopata, Van~Kuiken, Khalil, and
  Govind]{lopata2012linear}
Lopata,~K.; Van~Kuiken,~B.~E.; Khalil,~M.; Govind,~N. Linear-response and
  real-time time-dependent density functional theory studies of core-level
  near-edge x-ray absorption. \emph{J. Chem. Theo. Comput.} \textbf{2012},
  \emph{8}, 3284--3292\relax
\mciteBstWouldAddEndPuncttrue
\mciteSetBstMidEndSepPunct{\mcitedefaultmidpunct}
{\mcitedefaultendpunct}{\mcitedefaultseppunct}\relax
\EndOfBibitem
\bibitem[Besley(2021)]{besley2021modeling}
Besley,~N.~A. Modeling of the spectroscopy of core electrons with density
  functional theory. \emph{WIREs Comput. Mol. Sci.} \textbf{2021}, e1527\relax
\mciteBstWouldAddEndPuncttrue
\mciteSetBstMidEndSepPunct{\mcitedefaultmidpunct}
{\mcitedefaultendpunct}{\mcitedefaultseppunct}\relax
\EndOfBibitem
\bibitem[Zhang \latin{et~al.}(2012)Zhang, Biggs, Healion, Govind, and
  Mukamel]{zhang2012core}
Zhang,~Y.; Biggs,~J.~D.; Healion,~D.; Govind,~N.; Mukamel,~S. Core and valence
  excitations in resonant X-ray spectroscopy using restricted excitation window
  time-dependent density functional theory. \emph{J. Chem. Phys.}
  \textbf{2012}, \emph{137}, 194306\relax
\mciteBstWouldAddEndPuncttrue
\mciteSetBstMidEndSepPunct{\mcitedefaultmidpunct}
{\mcitedefaultendpunct}{\mcitedefaultseppunct}\relax
\EndOfBibitem
\bibitem[Besley(2020)]{besley2020density}
Besley,~N.~A. Density functional theory based methods for the calculation of
  X-ray spectroscopy. \emph{Acc. Chem. Res.} \textbf{2020}, \emph{53},
  1306--1315\relax
\mciteBstWouldAddEndPuncttrue
\mciteSetBstMidEndSepPunct{\mcitedefaultmidpunct}
{\mcitedefaultendpunct}{\mcitedefaultseppunct}\relax
\EndOfBibitem
\bibitem[Besley \latin{et~al.}(2009)Besley, Peach, and Tozer]{besley2009time}
Besley,~N.~A.; Peach,~M.~J.; Tozer,~D.~J. Time-dependent density functional
  theory calculations of near-edge X-ray absorption fine structure with
  short-range corrected functionals. \emph{Phys. Chem. Chem. Phys.}
  \textbf{2009}, \emph{11}, 10350--10358\relax
\mciteBstWouldAddEndPuncttrue
\mciteSetBstMidEndSepPunct{\mcitedefaultmidpunct}
{\mcitedefaultendpunct}{\mcitedefaultseppunct}\relax
\EndOfBibitem
\bibitem[Besley and Asmuruf(2010)Besley, and Asmuruf]{besley2010time}
Besley,~N.~A.; Asmuruf,~F.~A. Time-dependent density functional theory
  calculations of the spectroscopy of core electrons. \emph{Phys. Chem. Chem.
  Phys.} \textbf{2010}, \emph{12}, 12024--12039\relax
\mciteBstWouldAddEndPuncttrue
\mciteSetBstMidEndSepPunct{\mcitedefaultmidpunct}
{\mcitedefaultendpunct}{\mcitedefaultseppunct}\relax
\EndOfBibitem
\bibitem[Blake \latin{et~al.}(2018)Blake, Wei, Donahue, Lee, Keith, and
  Daly]{blake2018solid}
Blake,~A.~V.; Wei,~H.; Donahue,~C.~M.; Lee,~K.; Keith,~J.~M.; Daly,~S.~R. Solid
  energy calibration standards for P K-edge XANES: electronic structure
  analysis of \ce{PPh4Br}. \emph{J. Synchrotron Radiat.} \textbf{2018},
  \emph{25}, 529--536\relax
\mciteBstWouldAddEndPuncttrue
\mciteSetBstMidEndSepPunct{\mcitedefaultmidpunct}
{\mcitedefaultendpunct}{\mcitedefaultseppunct}\relax
\EndOfBibitem
\bibitem[Martin-Diaconescu and Kennepohl(2007)Martin-Diaconescu, and
  Kennepohl]{martin2007sulfur}
Martin-Diaconescu,~V.; Kennepohl,~P. Sulfur K-edge XAS as a probe of
  sulfur-centered radical intermediates. \emph{J. Am. Chem. Soc.}
  \textbf{2007}, \emph{129}, 3034--3035\relax
\mciteBstWouldAddEndPuncttrue
\mciteSetBstMidEndSepPunct{\mcitedefaultmidpunct}
{\mcitedefaultendpunct}{\mcitedefaultseppunct}\relax
\EndOfBibitem
\bibitem[Minasian \latin{et~al.}(2012)Minasian, Keith, Batista, Boland, Clark,
  Conradson, Kozimor, Martin, Schwarz, Shuh, \latin{et~al.}
  others]{minasian2012determining}
Minasian,~S.~G.; Keith,~J.~M.; Batista,~E.~R.; Boland,~K.~S.; Clark,~D.~L.;
  Conradson,~S.~D.; Kozimor,~S.~A.; Martin,~R.~L.; Schwarz,~D.~E.; Shuh,~D.~K.
  \latin{et~al.}  Determining relative f and d orbital contributions to M--Cl
  covalency in \ce{MCl6^2-}(M= Ti, Zr, Hf, U) and \ce{UOCl^5-}using Cl K-edge
  X-ray absorption spectroscopy and time-dependent density functional theory.
  \emph{J. Am. Chem. Soc.} \textbf{2012}, \emph{134}, 5586--5597\relax
\mciteBstWouldAddEndPuncttrue
\mciteSetBstMidEndSepPunct{\mcitedefaultmidpunct}
{\mcitedefaultendpunct}{\mcitedefaultseppunct}\relax
\EndOfBibitem
\bibitem[DeBeer~George \latin{et~al.}(2008)DeBeer~George, Petrenko, and
  Neese]{debeer2008prediction}
DeBeer~George,~S.; Petrenko,~T.; Neese,~F. Prediction of iron K-edge absorption
  spectra using time-dependent density functional theory. \emph{J. Phys. Chem.
  A} \textbf{2008}, \emph{112}, 12936--12943\relax
\mciteBstWouldAddEndPuncttrue
\mciteSetBstMidEndSepPunct{\mcitedefaultmidpunct}
{\mcitedefaultendpunct}{\mcitedefaultseppunct}\relax
\EndOfBibitem
\bibitem[Attar \latin{et~al.}(2017)Attar, Bhattacherjee, Pemmaraju, Schnorr,
  Closser, Prendergast, and Leone]{attar2017femtosecond}
Attar,~A.~R.; Bhattacherjee,~A.; Pemmaraju,~C.; Schnorr,~K.; Closser,~K.~D.;
  Prendergast,~D.; Leone,~S.~R. Femtosecond x-ray spectroscopy of an
  electrocyclic ring-opening reaction. \emph{Science} \textbf{2017},
  \emph{356}, 54--59\relax
\mciteBstWouldAddEndPuncttrue
\mciteSetBstMidEndSepPunct{\mcitedefaultmidpunct}
{\mcitedefaultendpunct}{\mcitedefaultseppunct}\relax
\EndOfBibitem
\bibitem[Stanton and Bartlett(1993)Stanton, and Bartlett]{stanton1993equation}
Stanton,~J.~F.; Bartlett,~R.~J. The equation of motion coupled-cluster method.
  A systematic biorthogonal approach to molecular excitation energies,
  transition probabilities, and excited state properties. \emph{J. Chem. Phys.}
  \textbf{1993}, \emph{98}, 7029--7039\relax
\mciteBstWouldAddEndPuncttrue
\mciteSetBstMidEndSepPunct{\mcitedefaultmidpunct}
{\mcitedefaultendpunct}{\mcitedefaultseppunct}\relax
\EndOfBibitem
\bibitem[Coriani and Koch(2015)Coriani, and Koch]{coriani2015communication}
Coriani,~S.; Koch,~H. Communication: X-ray absorption spectra and
  core-ionization potentials within a core-valence separated coupled cluster
  framework. \emph{J. Chem. Phys.} \textbf{2015}, \emph{143}, 181103\relax
\mciteBstWouldAddEndPuncttrue
\mciteSetBstMidEndSepPunct{\mcitedefaultmidpunct}
{\mcitedefaultendpunct}{\mcitedefaultseppunct}\relax
\EndOfBibitem
\bibitem[Peng \latin{et~al.}(2015)Peng, Lestrange, Goings, Caricato, and
  Li]{peng2015energy}
Peng,~B.; Lestrange,~P.~J.; Goings,~J.~J.; Caricato,~M.; Li,~X. Energy-specific
  equation-of-motion coupled-cluster methods for high-energy excited states:
  Application to K-edge X-ray absorption spectroscopy. \emph{J. Chem. Theo.
  Comput.} \textbf{2015}, \emph{11}, 4146--4153\relax
\mciteBstWouldAddEndPuncttrue
\mciteSetBstMidEndSepPunct{\mcitedefaultmidpunct}
{\mcitedefaultendpunct}{\mcitedefaultseppunct}\relax
\EndOfBibitem
\bibitem[Frati \latin{et~al.}(2019)Frati, De~Groot, Cerezo, Santoro, Cheng,
  Faber, and Coriani]{frati2019coupled}
Frati,~F.; De~Groot,~F.; Cerezo,~J.; Santoro,~F.; Cheng,~L.; Faber,~R.;
  Coriani,~S. Coupled cluster study of the x-ray absorption spectra of
  formaldehyde derivatives at the oxygen, carbon, and fluorine K-edges.
  \emph{J. Chem. Phys.} \textbf{2019}, \emph{151}, 064107\relax
\mciteBstWouldAddEndPuncttrue
\mciteSetBstMidEndSepPunct{\mcitedefaultmidpunct}
{\mcitedefaultendpunct}{\mcitedefaultseppunct}\relax
\EndOfBibitem
\bibitem[Vidal \latin{et~al.}(2019)Vidal, Feng, Epifanovsky, Krylov, and
  Coriani]{vidal2019new}
Vidal,~M.~L.; Feng,~X.; Epifanovsky,~E.; Krylov,~A.~I.; Coriani,~S. New and
  efficient equation-of-motion coupled-cluster framework for core-excited and
  core-ionized states. \emph{J. Chem. Theory Comput.} \textbf{2019}, \emph{15},
  3117--3133\relax
\mciteBstWouldAddEndPuncttrue
\mciteSetBstMidEndSepPunct{\mcitedefaultmidpunct}
{\mcitedefaultendpunct}{\mcitedefaultseppunct}\relax
\EndOfBibitem
\bibitem[Carbone \latin{et~al.}(2019)Carbone, Cheng, Myhre, Matthews, Koch, and
  Coriani]{carbone2019analysis}
Carbone,~J.~P.; Cheng,~L.; Myhre,~R.~H.; Matthews,~D.; Koch,~H.; Coriani,~S. An
  analysis of the performance of coupled cluster methods for K-edge core
  excitations and ionizations using standard basis sets. \emph{Adv. Quantum
  Chem.} \textbf{2019}, \emph{79}, 241--261\relax
\mciteBstWouldAddEndPuncttrue
\mciteSetBstMidEndSepPunct{\mcitedefaultmidpunct}
{\mcitedefaultendpunct}{\mcitedefaultseppunct}\relax
\EndOfBibitem
\bibitem[Besley \latin{et~al.}(2009)Besley, Gilbert, and Gill]{besley2009self}
Besley,~N.~A.; Gilbert,~A.~T.; Gill,~P. M.~W. Self-consistent-field
  calculations of core excited states. \emph{J. Chem. Phys.} \textbf{2009},
  \emph{130}, 124308\relax
\mciteBstWouldAddEndPuncttrue
\mciteSetBstMidEndSepPunct{\mcitedefaultmidpunct}
{\mcitedefaultendpunct}{\mcitedefaultseppunct}\relax
\EndOfBibitem
\bibitem[Hait and Head-Gordon(2021)Hait, and Head-Gordon]{hait2021orbital}
Hait,~D.; Head-Gordon,~M. Orbital optimized density functional theory for
  electronic excited states. \emph{J. Phys. Chem. Lett.} \textbf{2021},
  \emph{12}, 4517--4529\relax
\mciteBstWouldAddEndPuncttrue
\mciteSetBstMidEndSepPunct{\mcitedefaultmidpunct}
{\mcitedefaultendpunct}{\mcitedefaultseppunct}\relax
\EndOfBibitem
\bibitem[Derricotte and Evangelista(2015)Derricotte, and
  Evangelista]{derricotte2015simulation}
Derricotte,~W.~D.; Evangelista,~F.~A. Simulation of X-ray absorption spectra
  with orthogonality constrained density functional theory. \emph{Phys. Chem.
  Chem. Phys.} \textbf{2015}, \emph{17}, 14360--14374\relax
\mciteBstWouldAddEndPuncttrue
\mciteSetBstMidEndSepPunct{\mcitedefaultmidpunct}
{\mcitedefaultendpunct}{\mcitedefaultseppunct}\relax
\EndOfBibitem
\bibitem[Zheng \latin{et~al.}(2020)Zheng, Liu, Doumy, Young, and
  Cheng]{zheng2020hetero}
Zheng,~X.; Liu,~J.; Doumy,~G.; Young,~L.; Cheng,~L. Hetero-site Double Core
  Ionization Energies with Sub-electronvolt Accuracy from Delta-Coupled-Cluster
  Calculations. \emph{J. Phys. Chem. A} \textbf{2020}, \emph{124},
  4413--4426\relax
\mciteBstWouldAddEndPuncttrue
\mciteSetBstMidEndSepPunct{\mcitedefaultmidpunct}
{\mcitedefaultendpunct}{\mcitedefaultseppunct}\relax
\EndOfBibitem
\bibitem[Barca \latin{et~al.}(2018)Barca, Gilbert, and Gill]{barca2018simple}
Barca,~G.~M.; Gilbert,~A.~T.; Gill,~P. M.~W. Simple Models for Difficult
  Electronic Excitations. \emph{J. Chem. Theory Comput.} \textbf{2018},
  \emph{14}, 1501--1509\relax
\mciteBstWouldAddEndPuncttrue
\mciteSetBstMidEndSepPunct{\mcitedefaultmidpunct}
{\mcitedefaultendpunct}{\mcitedefaultseppunct}\relax
\EndOfBibitem
\bibitem[Shea \latin{et~al.}(2020)Shea, Gwin, and
  Neuscamman]{shea2020generalized}
Shea,~J.~A.; Gwin,~E.; Neuscamman,~E. A generalized variational principle with
  applications to excited state mean field theory. \emph{J. Chem. Theory
  Comput.} \textbf{2020}, \emph{16}, 1526--1540\relax
\mciteBstWouldAddEndPuncttrue
\mciteSetBstMidEndSepPunct{\mcitedefaultmidpunct}
{\mcitedefaultendpunct}{\mcitedefaultseppunct}\relax
\EndOfBibitem
\bibitem[Ye \latin{et~al.}(2017)Ye, Welborn, Ricke, and
  Van~Voorhis]{ye2017sigma}
Ye,~H.-Z.; Welborn,~M.; Ricke,~N.~D.; Van~Voorhis,~T. $\sigma$-SCF: A direct
  energy-targeting method to mean-field excited states. \emph{J. Chem. Phys.}
  \textbf{2017}, \emph{147}, 214104\relax
\mciteBstWouldAddEndPuncttrue
\mciteSetBstMidEndSepPunct{\mcitedefaultmidpunct}
{\mcitedefaultendpunct}{\mcitedefaultseppunct}\relax
\EndOfBibitem
\bibitem[Hait and Head-Gordon(2020)Hait, and Head-Gordon]{hait2020excited}
Hait,~D.; Head-Gordon,~M. Excited state orbital optimization via minimizing the
  square of the gradient: General approach and application to singly and doubly
  excited states via density functional theory. \emph{J. Chem. Theory Comput.}
  \textbf{2020}, \emph{16}, 1699--1710\relax
\mciteBstWouldAddEndPuncttrue
\mciteSetBstMidEndSepPunct{\mcitedefaultmidpunct}
{\mcitedefaultendpunct}{\mcitedefaultseppunct}\relax
\EndOfBibitem
\bibitem[Carter-Fenk and Herbert(2020)Carter-Fenk, and
  Herbert]{carter2020state}
Carter-Fenk,~K.; Herbert,~J.~M. State-Targeted Energy Projection: A Simple and
  Robust Approach to Orbital Relaxation of Non-Aufbau Self-Consistent Field
  Solutions. \emph{J. Chem. Theory Comput.} \textbf{2020}, \emph{16},
  5067--5082\relax
\mciteBstWouldAddEndPuncttrue
\mciteSetBstMidEndSepPunct{\mcitedefaultmidpunct}
{\mcitedefaultendpunct}{\mcitedefaultseppunct}\relax
\EndOfBibitem
\bibitem[Levi \latin{et~al.}(2020)Levi, Ivanov, and
  J{\'o}nsson]{levi2020variational}
Levi,~G.; Ivanov,~A.~V.; J{\'o}nsson,~H. Variational density functional
  calculations of excited states via direct optimization. \emph{J. Chem.
  Theory. Comput.} \textbf{2020}, \emph{16}, 6968--6982\relax
\mciteBstWouldAddEndPuncttrue
\mciteSetBstMidEndSepPunct{\mcitedefaultmidpunct}
{\mcitedefaultendpunct}{\mcitedefaultseppunct}\relax
\EndOfBibitem
\bibitem[Grofe \latin{et~al.}(2020)Grofe, Zhao, Wildman, Stetina, Li, Bao, and
  Gao]{grofe2020generalization}
Grofe,~A.; Zhao,~R.; Wildman,~A.; Stetina,~T.~F.; Li,~X.; Bao,~P.; Gao,~J.
  Generalization of Block-Localized Wave Function for Constrained Optimization
  of Excited Determinants. \emph{J. Chem. Theory Comput.} \textbf{2020},
  \emph{17}, 277--289\relax
\mciteBstWouldAddEndPuncttrue
\mciteSetBstMidEndSepPunct{\mcitedefaultmidpunct}
{\mcitedefaultendpunct}{\mcitedefaultseppunct}\relax
\EndOfBibitem
\bibitem[Hait and Head-Gordon(2020)Hait, and Head-Gordon]{hait2020highly}
Hait,~D.; Head-Gordon,~M. Highly Accurate Prediction of Core Spectra of
  Molecules at Density Functional Theory Cost: Attaining Sub-electronvolt Error
  from a Restricted Open-Shell Kohn--Sham Approach. \emph{J. Phys. Chem. Lett.}
  \textbf{2020}, \emph{11}, 775--786\relax
\mciteBstWouldAddEndPuncttrue
\mciteSetBstMidEndSepPunct{\mcitedefaultmidpunct}
{\mcitedefaultendpunct}{\mcitedefaultseppunct}\relax
\EndOfBibitem
\bibitem[Hait \latin{et~al.}(2020)Hait, Haugen, Yang, Oosterbaan, Leone, and
  Head-Gordon]{hait2020accurate}
Hait,~D.; Haugen,~E.~A.; Yang,~Z.; Oosterbaan,~K.~J.; Leone,~S.~R.;
  Head-Gordon,~M. Accurate prediction of core-level spectra of radicals at
  density functional theory cost via square gradient minimization and
  recoupling of mixed configurations. \emph{J. Chem. Phys.} \textbf{2020},
  \emph{153}, 134108\relax
\mciteBstWouldAddEndPuncttrue
\mciteSetBstMidEndSepPunct{\mcitedefaultmidpunct}
{\mcitedefaultendpunct}{\mcitedefaultseppunct}\relax
\EndOfBibitem
\bibitem[Garner and Neuscamman(2020)Garner, and Neuscamman]{garner2020core}
Garner,~S.~M.; Neuscamman,~E. Core excitations with excited state mean field
  and perturbation theory. \emph{J. Chem. Phys.} \textbf{2020}, \emph{153},
  154102\relax
\mciteBstWouldAddEndPuncttrue
\mciteSetBstMidEndSepPunct{\mcitedefaultmidpunct}
{\mcitedefaultendpunct}{\mcitedefaultseppunct}\relax
\EndOfBibitem
\bibitem[Zhao \latin{et~al.}(2021)Zhao, Grofe, Wang, Bao, Chen, Liu, and
  Gao]{zhao2021dynamic}
Zhao,~R.; Grofe,~A.; Wang,~Z.; Bao,~P.; Chen,~X.; Liu,~W.; Gao,~J.
  Dynamic-then-static approach for core excitations of open-shell molecules.
  \emph{J. Phys. Chem. Lett.} \textbf{2021}, \emph{12}, 7409--7417\relax
\mciteBstWouldAddEndPuncttrue
\mciteSetBstMidEndSepPunct{\mcitedefaultmidpunct}
{\mcitedefaultendpunct}{\mcitedefaultseppunct}\relax
\EndOfBibitem
\bibitem[Kahk \latin{et~al.}(2021)Kahk, Michelitsch, Maurer, Reuter, and
  Lischner]{kahk2021core}
Kahk,~J.~M.; Michelitsch,~G.~S.; Maurer,~R.~J.; Reuter,~K.; Lischner,~J. Core
  Electron Binding Energies in Solids from Periodic All-Electron
  $\Delta$-Self-Consistent-Field Calculations. \emph{J. Phys. Chem. Lett.}
  \textbf{2021}, \emph{12}, 9353--9359\relax
\mciteBstWouldAddEndPuncttrue
\mciteSetBstMidEndSepPunct{\mcitedefaultmidpunct}
{\mcitedefaultendpunct}{\mcitedefaultseppunct}\relax
\EndOfBibitem
\bibitem[Sun \latin{et~al.}(2015)Sun, Ruzsinszky, and Perdew]{SCAN}
Sun,~J.; Ruzsinszky,~A.; Perdew,~J.~P. {Strongly Constrained and Appropriately
  Normed Semilocal Density Functional}. \emph{Phys. Rev. Lett.} \textbf{2015},
  \emph{115}, 036402\relax
\mciteBstWouldAddEndPuncttrue
\mciteSetBstMidEndSepPunct{\mcitedefaultmidpunct}
{\mcitedefaultendpunct}{\mcitedefaultseppunct}\relax
\EndOfBibitem
\bibitem[Kahk and Lischner(2019)Kahk, and Lischner]{kahk2019accurate}
Kahk,~J.~M.; Lischner,~J. Accurate absolute core-electron binding energies of
  molecules, solids, and surfaces from first-principles calculations.
  \emph{Phys. Rev. Mat.} \textbf{2019}, \emph{3}, 100801\relax
\mciteBstWouldAddEndPuncttrue
\mciteSetBstMidEndSepPunct{\mcitedefaultmidpunct}
{\mcitedefaultendpunct}{\mcitedefaultseppunct}\relax
\EndOfBibitem
\bibitem[Takahashi(2017)]{takahashi2017relativistic}
Takahashi,~O. Relativistic corrections for single- and double-core excitation
  at the K-and L-edges from Li to Kr. \emph{Comput. Theor. Chem.}
  \textbf{2017}, \emph{1102}, 80--86\relax
\mciteBstWouldAddEndPuncttrue
\mciteSetBstMidEndSepPunct{\mcitedefaultmidpunct}
{\mcitedefaultendpunct}{\mcitedefaultseppunct}\relax
\EndOfBibitem
\bibitem[Norman and Dreuw(2018)Norman, and Dreuw]{norman2018simulating}
Norman,~P.; Dreuw,~A. Simulating X-ray spectroscopies and calculating
  core-excited states of molecules. \emph{Chem. Rev.} \textbf{2018},
  \emph{118}, 7208--7248\relax
\mciteBstWouldAddEndPuncttrue
\mciteSetBstMidEndSepPunct{\mcitedefaultmidpunct}
{\mcitedefaultendpunct}{\mcitedefaultseppunct}\relax
\EndOfBibitem
\bibitem[Bussy and Hutter(2021)Bussy, and Hutter]{bussy2021efficient}
Bussy,~A.; Hutter,~J. Efficient and low-scaling linear-response time-dependent
  density functional theory implementation for core-level spectroscopy of large
  and periodic systems. \emph{Phys. Chem. Chem. Phys.} \textbf{2021},
  \emph{23}, 4736--4746\relax
\mciteBstWouldAddEndPuncttrue
\mciteSetBstMidEndSepPunct{\mcitedefaultmidpunct}
{\mcitedefaultendpunct}{\mcitedefaultseppunct}\relax
\EndOfBibitem
\bibitem[Stetina \latin{et~al.}(2019)Stetina, Kasper, and
  Li]{stetina2019modeling}
Stetina,~T.~F.; Kasper,~J.~M.; Li,~X. Modeling L2, 3-edge X-ray absorption
  spectroscopy with linear response exact two-component relativistic
  time-dependent density functional theory. \emph{J. Chem. Phys.}
  \textbf{2019}, \emph{150}, 234103\relax
\mciteBstWouldAddEndPuncttrue
\mciteSetBstMidEndSepPunct{\mcitedefaultmidpunct}
{\mcitedefaultendpunct}{\mcitedefaultseppunct}\relax
\EndOfBibitem
\bibitem[Repisky \latin{et~al.}(2015)Repisky, Konecny, Kadek, Komorovsky,
  Malkin, Malkin, and Ruud]{repisky2015excitation}
Repisky,~M.; Konecny,~L.; Kadek,~M.; Komorovsky,~S.; Malkin,~O.~L.;
  Malkin,~V.~G.; Ruud,~K. Excitation energies from real-time propagation of the
  four-component Dirac--Kohn--Sham equation. \emph{J. Chem. Theory Comput.}
  \textbf{2015}, \emph{11}, 980--991\relax
\mciteBstWouldAddEndPuncttrue
\mciteSetBstMidEndSepPunct{\mcitedefaultmidpunct}
{\mcitedefaultendpunct}{\mcitedefaultseppunct}\relax
\EndOfBibitem
\bibitem[Liu and Cheng(2021)Liu, and Cheng]{liu2021relativistic}
Liu,~J.; Cheng,~L. Relativistic coupled-cluster and equation-of-motion
  coupled-cluster methods. \emph{WIREs Comput. Mol. Sci.} \textbf{2021},
  e1536\relax
\mciteBstWouldAddEndPuncttrue
\mciteSetBstMidEndSepPunct{\mcitedefaultmidpunct}
{\mcitedefaultendpunct}{\mcitedefaultseppunct}\relax
\EndOfBibitem
\bibitem[Halbert \latin{et~al.}(2021)Halbert, Vidal, Shee, Coriani, and Severo
  Pereira~Gomes]{halbert2021relativistic}
Halbert,~L.; Vidal,~M.~L.; Shee,~A.; Coriani,~S.; Severo Pereira~Gomes,~A.
  Relativistic EOM-CCSD for Core-Excited and Core-Ionized State Energies Based
  on the Four-Component Dirac--Coulomb (- Gaunt) Hamiltonian. \emph{J. Chem.
  Theory Comput.} \textbf{2021}, \emph{17}, 3583--3598\relax
\mciteBstWouldAddEndPuncttrue
\mciteSetBstMidEndSepPunct{\mcitedefaultmidpunct}
{\mcitedefaultendpunct}{\mcitedefaultseppunct}\relax
\EndOfBibitem
\bibitem[Dyall(1997)]{dyall1997interfacing}
Dyall,~K.~G. Interfacing relativistic and nonrelativistic methods. I.
  Normalized elimination of the small component in the modified Dirac equation.
  \emph{J. Chem. Phys.} \textbf{1997}, \emph{106}, 9618--9626\relax
\mciteBstWouldAddEndPuncttrue
\mciteSetBstMidEndSepPunct{\mcitedefaultmidpunct}
{\mcitedefaultendpunct}{\mcitedefaultseppunct}\relax
\EndOfBibitem
\bibitem[Kutzelnigg and Liu(2005)Kutzelnigg, and
  Liu]{kutzelnigg2005quasirelativistic}
Kutzelnigg,~W.; Liu,~W. Quasirelativistic theory equivalent to fully
  relativistic theory. \emph{J. Chem. Phys.} \textbf{2005}, \emph{123},
  241102\relax
\mciteBstWouldAddEndPuncttrue
\mciteSetBstMidEndSepPunct{\mcitedefaultmidpunct}
{\mcitedefaultendpunct}{\mcitedefaultseppunct}\relax
\EndOfBibitem
\bibitem[Ilias and Saue(2007)Ilias, and Saue]{saue2007inf}
Ilias,~M.; Saue,~T. An Infinite-Order Relativistic Hamiltonian by a Simple
  One-Step Transformation. \emph{J. Chem. Phys.} \textbf{2007}, \emph{126},
  064102\relax
\mciteBstWouldAddEndPuncttrue
\mciteSetBstMidEndSepPunct{\mcitedefaultmidpunct}
{\mcitedefaultendpunct}{\mcitedefaultseppunct}\relax
\EndOfBibitem
\bibitem[Liu and Peng(2009)Liu, and Peng]{Liu2009x2c}
Liu,~W.; Peng,~D. Exact Two-component Hamiltonians Revisited. \emph{J. Chem.
  Phys.} \textbf{2009}, \emph{131}, 031104\relax
\mciteBstWouldAddEndPuncttrue
\mciteSetBstMidEndSepPunct{\mcitedefaultmidpunct}
{\mcitedefaultendpunct}{\mcitedefaultseppunct}\relax
\EndOfBibitem
\bibitem[Saue(2011)]{saue2011primer}
Saue,~T. Relativistic Hamiltonians for Chemistry: A Primer. \emph{Chem. Phys.
  Chem.} \textbf{2011}, \emph{12}, 3077--3094\relax
\mciteBstWouldAddEndPuncttrue
\mciteSetBstMidEndSepPunct{\mcitedefaultmidpunct}
{\mcitedefaultendpunct}{\mcitedefaultseppunct}\relax
\EndOfBibitem
\bibitem[Li \latin{et~al.}(2012)Li, Xiao, and Liu]{Liu2012spin}
Li,~Z.; Xiao,~Y.; Liu,~W. On the spin separation of algebraic two-component
  relativistic Hamiltonians. \emph{J. Chem. Phys.} \textbf{2012}, \emph{137},
  154114\relax
\mciteBstWouldAddEndPuncttrue
\mciteSetBstMidEndSepPunct{\mcitedefaultmidpunct}
{\mcitedefaultendpunct}{\mcitedefaultseppunct}\relax
\EndOfBibitem
\bibitem[Cheng and Gauss(2011)Cheng, and Gauss]{cheng2011analytic}
Cheng,~L.; Gauss,~J. Analytic energy gradients for the spin-free exact
  two-component theory using an exact block diagonalization for the
  one-electron Dirac Hamiltonian. \emph{J. Chem. Phys.} \textbf{2011},
  \emph{135}, 084114\relax
\mciteBstWouldAddEndPuncttrue
\mciteSetBstMidEndSepPunct{\mcitedefaultmidpunct}
{\mcitedefaultendpunct}{\mcitedefaultseppunct}\relax
\EndOfBibitem
\bibitem[Verma \latin{et~al.}(2016)Verma, Derricotte, and
  Evangelista]{verma2016predicting}
Verma,~P.; Derricotte,~W.~D.; Evangelista,~F.~A. Predicting near edge X-ray
  absorption spectra with the spin-free exact-two-component Hamiltonian and
  orthogonality constrained density functional theory. \emph{J. Chem. Theory
  Comput.} \textbf{2016}, \emph{12}, 144--156\relax
\mciteBstWouldAddEndPuncttrue
\mciteSetBstMidEndSepPunct{\mcitedefaultmidpunct}
{\mcitedefaultendpunct}{\mcitedefaultseppunct}\relax
\EndOfBibitem
\bibitem[Kohn and Sham(1965)Kohn, and Sham]{kohn1965self}
Kohn,~W.; Sham,~L.~J. {Self-consistent equations including exchange and
  correlation effects}. \emph{Phys. Rev.} \textbf{1965}, \emph{140},
  A1133--A1138\relax
\mciteBstWouldAddEndPuncttrue
\mciteSetBstMidEndSepPunct{\mcitedefaultmidpunct}
{\mcitedefaultendpunct}{\mcitedefaultseppunct}\relax
\EndOfBibitem
\bibitem[Bagus(1965)]{bagus1965self}
Bagus,~P.~S. Self-consistent-field wave functions for hole states of some
  Ne-like and Ar-like ions. \emph{Phys. Rev.} \textbf{1965}, \emph{139},
  A619--A634\relax
\mciteBstWouldAddEndPuncttrue
\mciteSetBstMidEndSepPunct{\mcitedefaultmidpunct}
{\mcitedefaultendpunct}{\mcitedefaultseppunct}\relax
\EndOfBibitem
\bibitem[Ziegler \latin{et~al.}(1977)Ziegler, Rauk, and
  Baerends]{ziegler1977calculation}
Ziegler,~T.; Rauk,~A.; Baerends,~E.~J. On the calculation of multiplet energies
  by the Hartree-Fock-Slater method. \emph{Theor. Chim. Acta.} \textbf{1977},
  \emph{43}, 261--271\relax
\mciteBstWouldAddEndPuncttrue
\mciteSetBstMidEndSepPunct{\mcitedefaultmidpunct}
{\mcitedefaultendpunct}{\mcitedefaultseppunct}\relax
\EndOfBibitem
\bibitem[Niskanen \latin{et~al.}(2011)Niskanen, Norman, Aksela, and
  {\AA}gren]{niskanen2011relativistic}
Niskanen,~J.; Norman,~P.; Aksela,~H.; {\AA}gren,~H. Relativistic contributions
  to single and double core electron ionization energies of noble gases.
  \emph{J. Chem. Phys.} \textbf{2011}, \emph{135}, 054310\relax
\mciteBstWouldAddEndPuncttrue
\mciteSetBstMidEndSepPunct{\mcitedefaultmidpunct}
{\mcitedefaultendpunct}{\mcitedefaultseppunct}\relax
\EndOfBibitem
\bibitem[Zheng and Cheng(2019)Zheng, and Cheng]{zheng2019performance}
Zheng,~X.; Cheng,~L. Performance of Delta-Coupled-Cluster Methods for
  Calculations of Core-Ionization Energies of First-Row Elements. \emph{J.
  Chem. Theory Comput.} \textbf{2019}, \emph{15}, 4945--4955\relax
\mciteBstWouldAddEndPuncttrue
\mciteSetBstMidEndSepPunct{\mcitedefaultmidpunct}
{\mcitedefaultendpunct}{\mcitedefaultseppunct}\relax
\EndOfBibitem
\bibitem[Frank \latin{et~al.}(1998)Frank, Hutter, Marx, and
  Parrinello]{frank1998molecular}
Frank,~I.; Hutter,~J.; Marx,~D.; Parrinello,~M. Molecular dynamics in low-spin
  excited states. \emph{J. Chem. Phys.} \textbf{1998}, \emph{108},
  4060--4069\relax
\mciteBstWouldAddEndPuncttrue
\mciteSetBstMidEndSepPunct{\mcitedefaultmidpunct}
{\mcitedefaultendpunct}{\mcitedefaultseppunct}\relax
\EndOfBibitem
\bibitem[Kowalczyk \latin{et~al.}(2013)Kowalczyk, Tsuchimochi, Chen, Top, and
  Van~Voorhis]{kowalczyk2013excitation}
Kowalczyk,~T.; Tsuchimochi,~T.; Chen,~P.-T.; Top,~L.; Van~Voorhis,~T.
  Excitation energies and Stokes shifts from a restricted open-shell Kohn-Sham
  approach. \emph{J. Chem. Phys.} \textbf{2013}, \emph{138}, 164101\relax
\mciteBstWouldAddEndPuncttrue
\mciteSetBstMidEndSepPunct{\mcitedefaultmidpunct}
{\mcitedefaultendpunct}{\mcitedefaultseppunct}\relax
\EndOfBibitem
\bibitem[{\AA}gren \latin{et~al.}(1978){\AA}gren, Nordgren, Selander, Nordling,
  and Siegbahn]{aagren1978multiplet}
{\AA}gren,~H.; Nordgren,~J.; Selander,~L.; Nordling,~C.; Siegbahn,~K. Multiplet
  structure in the high-resolution x-ray emission spectrum of neon. \emph{J.
  Electron Spectrosc. Relat. Phenom.} \textbf{1978}, \emph{14}, 27--39\relax
\mciteBstWouldAddEndPuncttrue
\mciteSetBstMidEndSepPunct{\mcitedefaultmidpunct}
{\mcitedefaultendpunct}{\mcitedefaultseppunct}\relax
\EndOfBibitem
\bibitem[Banna \latin{et~al.}(1978)Banna, Wallbank, Frost, McDowell, and
  Perera]{banna1978free}
Banna,~M.; Wallbank,~B.; Frost,~D.; McDowell,~C.; Perera,~J. Free atom core
  binding energies from X-ray photoelectron spectroscopy. II. Na, K, Rb, Cs,
  and Mg. \emph{J. Chem. Phys.} \textbf{1978}, \emph{68}, 5459--5466\relax
\mciteBstWouldAddEndPuncttrue
\mciteSetBstMidEndSepPunct{\mcitedefaultmidpunct}
{\mcitedefaultendpunct}{\mcitedefaultseppunct}\relax
\EndOfBibitem
\bibitem[Bodeur \latin{et~al.}(1990)Bodeur, Milli{\'e}, and
  Nenner]{bodeur1990single}
Bodeur,~S.; Milli{\'e},~P.; Nenner,~I. Single-and multiple-electron effects in
  the Si 1s photoabsorption spectra of \ce{SiX4} (X= H, D, F, Cl, Br, \ce{CH3},
  \ce{C2H5}, \ce{OCH3}, \ce{OC2H5}) molecules: Experiment and theory.
  \emph{Phys. Rev. A} \textbf{1990}, \emph{41}, 252--263\relax
\mciteBstWouldAddEndPuncttrue
\mciteSetBstMidEndSepPunct{\mcitedefaultmidpunct}
{\mcitedefaultendpunct}{\mcitedefaultseppunct}\relax
\EndOfBibitem
\bibitem[Sodhi and Cavell(1983)Sodhi, and Cavell]{sodhi1983kll}
Sodhi,~R.~N.; Cavell,~R.~G. KLL Auger and core-level (1s and 2p) photoelectron
  shifts in a series of gaseous phosphorus compounds. \emph{J. Electron
  Spectrosc. Relat. Phenom.} \textbf{1983}, \emph{32}, 283--312\relax
\mciteBstWouldAddEndPuncttrue
\mciteSetBstMidEndSepPunct{\mcitedefaultmidpunct}
{\mcitedefaultendpunct}{\mcitedefaultseppunct}\relax
\EndOfBibitem
\bibitem[Keski-Rahkonen and Krause(1976)Keski-Rahkonen, and
  Krause]{keski1976energies}
Keski-Rahkonen,~O.; Krause,~M. Energies and chemical shifts of the sulphur 1s
  level and the KL2L3 ($^1$D$_2$) Auger line in \ce{H2S}, \ce{SO2} and
  \ce{SF6}. \emph{J. Electron Spectrosc. Relat. Phenom.} \textbf{1976},
  \emph{9}, 371--380\relax
\mciteBstWouldAddEndPuncttrue
\mciteSetBstMidEndSepPunct{\mcitedefaultmidpunct}
{\mcitedefaultendpunct}{\mcitedefaultseppunct}\relax
\EndOfBibitem
\bibitem[Perera and LaVilla(1984)Perera, and LaVilla]{perera1984molecular}
Perera,~R.~C.; LaVilla,~R.~E. Molecular x-ray spectra: S-K $\beta$ emission and
  K absorption spectra of \ce{SCO} and \ce{CS2}. \emph{J. Chem. Phys.}
  \textbf{1984}, \emph{81}, 3375--3382\relax
\mciteBstWouldAddEndPuncttrue
\mciteSetBstMidEndSepPunct{\mcitedefaultmidpunct}
{\mcitedefaultendpunct}{\mcitedefaultseppunct}\relax
\EndOfBibitem
\bibitem[Sodhi and Cavell(1986)Sodhi, and Cavell]{sodhi1986kll}
Sodhi,~R.~N.; Cavell,~R.~G. KLL auger and core level (1s and 2p) photoelectron
  shifts in a series of gaseous sulfur compounds. \emph{J. Electron Spectrosc.
  Relat. Phenom.} \textbf{1986}, \emph{41}, 1--24\relax
\mciteBstWouldAddEndPuncttrue
\mciteSetBstMidEndSepPunct{\mcitedefaultmidpunct}
{\mcitedefaultendpunct}{\mcitedefaultseppunct}\relax
\EndOfBibitem
\bibitem[Bodeur \latin{et~al.}(1990)Bodeur, Mar{\'e}chal, Reynaud, Bazin, and
  Nenner]{bodeur1990chlorine}
Bodeur,~S.; Mar{\'e}chal,~J.; Reynaud,~C.; Bazin,~D.; Nenner,~I. Chlorine K
  shell photoabsorption spectra of gas phase \ce{HCl} and \ce{Cl2} molecules.
  \emph{Zeitschrift f{\"u}r Physik D Atoms, Molecules and Clusters}
  \textbf{1990}, \emph{17}, 291--298\relax
\mciteBstWouldAddEndPuncttrue
\mciteSetBstMidEndSepPunct{\mcitedefaultmidpunct}
{\mcitedefaultendpunct}{\mcitedefaultseppunct}\relax
\EndOfBibitem
\bibitem[Lindle \latin{et~al.}(1991)Lindle, Cowan, Jach, LaVilla, Deslattes,
  and Perera]{lindle1991polarized}
Lindle,~D.~W.; Cowan,~P.; Jach,~T.; LaVilla,~R.; Deslattes,~R.; Perera,~R.~C.
  Polarized x-ray emission studies of methyl chloride and the
  chlorofluoromethanes. \emph{Phys. Rev. A} \textbf{1991}, \emph{43},
  2353--2366\relax
\mciteBstWouldAddEndPuncttrue
\mciteSetBstMidEndSepPunct{\mcitedefaultmidpunct}
{\mcitedefaultendpunct}{\mcitedefaultseppunct}\relax
\EndOfBibitem
\bibitem[Reynaud \latin{et~al.}(1992)Reynaud, Bodeur, Mar{\'e}chal, Bazin,
  Milli{\'e}, Nenner, Rockland, and Baumg{\"a}rtel]{reynaud1992electronic}
Reynaud,~C.; Bodeur,~S.; Mar{\'e}chal,~J.; Bazin,~D.; Milli{\'e},~P.;
  Nenner,~I.; Rockland,~U.; Baumg{\"a}rtel,~H. Electronic properties of the
  \ce{SF5Cl} molecule: a comparison with \ce{SF6}. I. Photoabsorption spectra
  near the sulphur K and chlorine K edges. \emph{Chem. Phys.} \textbf{1992},
  \emph{166}, 411--424\relax
\mciteBstWouldAddEndPuncttrue
\mciteSetBstMidEndSepPunct{\mcitedefaultmidpunct}
{\mcitedefaultendpunct}{\mcitedefaultseppunct}\relax
\EndOfBibitem
\bibitem[Breinig \latin{et~al.}(1980)Breinig, Chen, Ice, Parente, Crasemann,
  and Brown]{breinig1980atomic}
Breinig,~M.; Chen,~M.~H.; Ice,~G.~E.; Parente,~F.; Crasemann,~B.; Brown,~G.~S.
  Atomic inner-shell level energies determined by absorption spectrometry with
  synchrotron radiation. \emph{Phys. Rev. A} \textbf{1980}, \emph{22},
  520--528\relax
\mciteBstWouldAddEndPuncttrue
\mciteSetBstMidEndSepPunct{\mcitedefaultmidpunct}
{\mcitedefaultendpunct}{\mcitedefaultseppunct}\relax
\EndOfBibitem
\bibitem[Ambroise and Jensen(2018)Ambroise, and Jensen]{ambroise2018probing}
Ambroise,~M.~A.; Jensen,~F. Probing Basis Set Requirements for Calculating Core
  Ionization and Core Excitation Spectroscopy by the $\Delta$
  Self-Consistent-Field Approach. \emph{J. Chem. Theory Comput.} \textbf{2018},
  \emph{15}, 325--337\relax
\mciteBstWouldAddEndPuncttrue
\mciteSetBstMidEndSepPunct{\mcitedefaultmidpunct}
{\mcitedefaultendpunct}{\mcitedefaultseppunct}\relax
\EndOfBibitem
\bibitem[Jensen(2014)]{jensen2014unifying}
Jensen,~F. Unifying general and segmented contracted basis sets. Segmented
  polarization consistent basis sets. \emph{J. Chem. Theory Comput.}
  \textbf{2014}, \emph{10}, 1074--1085\relax
\mciteBstWouldAddEndPuncttrue
\mciteSetBstMidEndSepPunct{\mcitedefaultmidpunct}
{\mcitedefaultendpunct}{\mcitedefaultseppunct}\relax
\EndOfBibitem
\bibitem[Becke(1993)]{b3lyp}
Becke,~A.~D. {Density-functional thermochemistry. III. The role of exact
  exchange}. \emph{J. Chem. Phys.} \textbf{1993}, \emph{98}, 5648--5652\relax
\mciteBstWouldAddEndPuncttrue
\mciteSetBstMidEndSepPunct{\mcitedefaultmidpunct}
{\mcitedefaultendpunct}{\mcitedefaultseppunct}\relax
\EndOfBibitem
\bibitem[Stephens \latin{et~al.}(1994)Stephens, Devlin, Chabalowski, and
  Frisch]{stephens1994ab}
Stephens,~P.~J.; Devlin,~F.~J.; Chabalowski,~C.~F.; Frisch,~M.~J. Ab initio
  calculation of vibrational absorption and circular dichroism spectra using
  density functional force fields. \emph{J. Phys. Chem.} \textbf{1994},
  \emph{98}, 11623--11627\relax
\mciteBstWouldAddEndPuncttrue
\mciteSetBstMidEndSepPunct{\mcitedefaultmidpunct}
{\mcitedefaultendpunct}{\mcitedefaultseppunct}\relax
\EndOfBibitem
\bibitem[Adamo and Barone(1999)Adamo, and Barone]{pbe0}
Adamo,~C.; Barone,~V. {Toward reliable density functional methods without
  adjustable parameters: The PBE0 model}. \emph{J. Chem. Phys.} \textbf{1999},
  \emph{110}, 6158--6170\relax
\mciteBstWouldAddEndPuncttrue
\mciteSetBstMidEndSepPunct{\mcitedefaultmidpunct}
{\mcitedefaultendpunct}{\mcitedefaultseppunct}\relax
\EndOfBibitem
\bibitem[Tao \latin{et~al.}(2003)Tao, Perdew, Staroverov, and Scuseria]{tpss}
Tao,~J.; Perdew,~J.~P.; Staroverov,~V.~N.; Scuseria,~G.~E. {Climbing the
  density functional ladder: Nonempirical meta--generalized gradient
  approximation designed for molecules and solids}. \emph{Phys. Rev. Lett.}
  \textbf{2003}, \emph{91}, 146401\relax
\mciteBstWouldAddEndPuncttrue
\mciteSetBstMidEndSepPunct{\mcitedefaultmidpunct}
{\mcitedefaultendpunct}{\mcitedefaultseppunct}\relax
\EndOfBibitem
\bibitem[Chan(2021)]{chan2021assessment}
Chan,~B. Assessment and development of DFT with the expanded CUAGAU-2 set of
  group-11 cluster systems. \emph{Int. J. Quantum Chem.} \textbf{2021},
  \emph{121}, e26453\relax
\mciteBstWouldAddEndPuncttrue
\mciteSetBstMidEndSepPunct{\mcitedefaultmidpunct}
{\mcitedefaultendpunct}{\mcitedefaultseppunct}\relax
\EndOfBibitem
\bibitem[Hui and Chai(2016)Hui, and Chai]{scan0}
Hui,~K.; Chai,~J.-D. {SCAN-based hybrid and double-hybrid density functionals
  from models without fitted parameters}. \emph{J. Chem. Phys.} \textbf{2016},
  \emph{144}, 044114\relax
\mciteBstWouldAddEndPuncttrue
\mciteSetBstMidEndSepPunct{\mcitedefaultmidpunct}
{\mcitedefaultendpunct}{\mcitedefaultseppunct}\relax
\EndOfBibitem
\bibitem[Becke(1993)]{bhhlyp}
Becke,~A.~D. A new mixing of Hartree--Fock and local density-functional
  theories. \emph{J. Chem. Phys.} \textbf{1993}, \emph{98}, 1372--1377\relax
\mciteBstWouldAddEndPuncttrue
\mciteSetBstMidEndSepPunct{\mcitedefaultmidpunct}
{\mcitedefaultendpunct}{\mcitedefaultseppunct}\relax
\EndOfBibitem
\bibitem[Perdew \latin{et~al.}(1996)Perdew, Burke, and Ernzerhof]{PBE}
Perdew,~J.~P.; Burke,~K.; Ernzerhof,~M. {Generalized gradient approximation
  made simple}. \emph{Phys. Rev. Lett.} \textbf{1996}, \emph{77},
  3865--3868\relax
\mciteBstWouldAddEndPuncttrue
\mciteSetBstMidEndSepPunct{\mcitedefaultmidpunct}
{\mcitedefaultendpunct}{\mcitedefaultseppunct}\relax
\EndOfBibitem
\bibitem[Cavell and J{\"u}rgensen(1999)Cavell, and
  J{\"u}rgensen]{cavell1999chemical}
Cavell,~R.~G.; J{\"u}rgensen,~A. Chemical shifts in P-1s photoabsorption
  spectra of gaseous phosphorus compounds. \emph{J. Electron Spectrosc. Relat.
  Phenom.} \textbf{1999}, \emph{101}, 125--129\relax
\mciteBstWouldAddEndPuncttrue
\mciteSetBstMidEndSepPunct{\mcitedefaultmidpunct}
{\mcitedefaultendpunct}{\mcitedefaultseppunct}\relax
\EndOfBibitem
\bibitem[Reynaud \latin{et~al.}(1996)Reynaud, Gaveau, Bisson, Milli{\'e},
  Nenner, Bodeur, Archirel, and L{\'e}vy]{reynaud1996double}
Reynaud,~C.; Gaveau,~M.-A.; Bisson,~K.; Milli{\'e},~P.; Nenner,~I.; Bodeur,~S.;
  Archirel,~P.; L{\'e}vy,~B. Double-core ionization and excitation above the
  sulphur K-edge in \ce{H2S}, \ce{SO2} and \ce{SF6}. \emph{J. Phys. B: At. Mol.
  Opt. Phys.} \textbf{1996}, \emph{29}, 5403--5419\relax
\mciteBstWouldAddEndPuncttrue
\mciteSetBstMidEndSepPunct{\mcitedefaultmidpunct}
{\mcitedefaultendpunct}{\mcitedefaultseppunct}\relax
\EndOfBibitem
\bibitem[Bodeur and Hitchcock(1987)Bodeur, and Hitchcock]{bodeur1987inner}
Bodeur,~S.; Hitchcock,~A. Inner-and valence-shell excitation of \ce{SF4}
  studied by photoabsorption and electron energy loss spectroscopy. \emph{Chem.
  Phys.} \textbf{1987}, \emph{111}, 467--479\relax
\mciteBstWouldAddEndPuncttrue
\mciteSetBstMidEndSepPunct{\mcitedefaultmidpunct}
{\mcitedefaultendpunct}{\mcitedefaultseppunct}\relax
\EndOfBibitem
\bibitem[Evangelista \latin{et~al.}(2013)Evangelista, Shushkov, and
  Tully]{evangelista2013orthogonality}
Evangelista,~F.~A.; Shushkov,~P.; Tully,~J.~C. Orthogonality constrained
  density functional theory for electronic excited states. \emph{J. Phys. Chem.
  A} \textbf{2013}, \emph{117}, 7378--7392\relax
\mciteBstWouldAddEndPuncttrue
\mciteSetBstMidEndSepPunct{\mcitedefaultmidpunct}
{\mcitedefaultendpunct}{\mcitedefaultseppunct}\relax
\EndOfBibitem
\bibitem[Woon and Dunning~Jr(1995)Woon, and Dunning~Jr]{woon1995gaussian}
Woon,~D.~E.; Dunning~Jr,~T.~H. {Gaussian basis sets for use in correlated
  molecular calculations. V. Core-valence basis sets for boron through neon}.
  \emph{J. Chem. Phys.} \textbf{1995}, \emph{103}, 4572--4585\relax
\mciteBstWouldAddEndPuncttrue
\mciteSetBstMidEndSepPunct{\mcitedefaultmidpunct}
{\mcitedefaultendpunct}{\mcitedefaultseppunct}\relax
\EndOfBibitem
\bibitem[Mennucci(2012)]{mennucci2012polarizable}
Mennucci,~B. Polarizable continuum model. \emph{Wiley Interdiscip. Rev. Comput.
  Mol. Sci.} \textbf{2012}, \emph{2}, 386--404\relax
\mciteBstWouldAddEndPuncttrue
\mciteSetBstMidEndSepPunct{\mcitedefaultmidpunct}
{\mcitedefaultendpunct}{\mcitedefaultseppunct}\relax
\EndOfBibitem
\bibitem[Cances \latin{et~al.}(1997)Cances, Mennucci, and
  Tomasi]{cances1997new}
Cances,~E.; Mennucci,~B.; Tomasi,~J. A new integral equation formalism for the
  polarizable continuum model: Theoretical background and applications to
  isotropic and anisotropic dielectrics. \emph{J. Chem. Phys.} \textbf{1997},
  \emph{107}, 3032--3041\relax
\mciteBstWouldAddEndPuncttrue
\mciteSetBstMidEndSepPunct{\mcitedefaultmidpunct}
{\mcitedefaultendpunct}{\mcitedefaultseppunct}\relax
\EndOfBibitem
\bibitem[Kunze \latin{et~al.}(2021)Kunze, Hansen, Grimme, and
  Mewes]{kunze2021pcm}
Kunze,~L.; Hansen,~A.; Grimme,~S.; Mewes,~J.-M. PCM-ROKS for the Description of
  Charge-Transfer States in Solution: Singlet--Triplet Gaps with Chemical
  Accuracy from Open-Shell Kohn--Sham Reaction-Field Calculations. \emph{J.
  Phys. Chem. Lett.} \textbf{2021}, \emph{12}, 8470--8480\relax
\mciteBstWouldAddEndPuncttrue
\mciteSetBstMidEndSepPunct{\mcitedefaultmidpunct}
{\mcitedefaultendpunct}{\mcitedefaultseppunct}\relax
\EndOfBibitem
\bibitem[Sutherland \latin{et~al.}(1993)Sutherland, Kasrai, Bancroft, Liu, and
  Tan]{sutherland1993si}
Sutherland,~D.; Kasrai,~M.; Bancroft,~G.; Liu,~Z.; Tan,~K. Si L-and K-edge
  x-ray-absorption near-edge spectroscopy of gas-phase
  \ce{Si(CH3)_x(OCH3)_{4-x}}: Models for solid-state analogs. \emph{Phys. Rev.
  B} \textbf{1993}, \emph{48}, 14989--15001\relax
\mciteBstWouldAddEndPuncttrue
\mciteSetBstMidEndSepPunct{\mcitedefaultmidpunct}
{\mcitedefaultendpunct}{\mcitedefaultseppunct}\relax
\EndOfBibitem
\bibitem[Engemann \latin{et~al.}(1997)Engemann, Kohring, Pantelouris, Hormes,
  Grimme, Peyerimhoff, Clade, Frick, and Jansen]{engemann1997experimental}
Engemann,~C.; Kohring,~G.; Pantelouris,~A.; Hormes,~J.; Grimme,~S.;
  Peyerimhoff,~S.; Clade,~J.; Frick,~F.; Jansen,~M. Experimental and
  theoretical investigations of the X-ray absorption near edge spectra (XANES)
  of \ce{P4O6} and \ce{P4O6X} (X = O, S, Se). \emph{Chem. Phys.} \textbf{1997},
  \emph{221}, 189--198\relax
\mciteBstWouldAddEndPuncttrue
\mciteSetBstMidEndSepPunct{\mcitedefaultmidpunct}
{\mcitedefaultendpunct}{\mcitedefaultseppunct}\relax
\EndOfBibitem
\bibitem[Ibuki \latin{et~al.}(2004)Ibuki, Shimada, Nagaoka, Fujii, Hino,
  Kakiuchi, Okada, Tabayashi, Matsudo, Yamana, \latin{et~al.}
  others]{ibuki2004total}
Ibuki,~T.; Shimada,~Y.; Nagaoka,~S.; Fujii,~A.; Hino,~M.; Kakiuchi,~T.;
  Okada,~K.; Tabayashi,~K.; Matsudo,~T.; Yamana,~Y. \latin{et~al.}  Total
  photoabsorption cross-sections of \ce{CF3SF5} in the C, F and S K-shell
  regions. \emph{Chem. Phys. Lett.} \textbf{2004}, \emph{392}, 303--308\relax
\mciteBstWouldAddEndPuncttrue
\mciteSetBstMidEndSepPunct{\mcitedefaultmidpunct}
{\mcitedefaultendpunct}{\mcitedefaultseppunct}\relax
\EndOfBibitem
\bibitem[Ochmann \latin{et~al.}(2018)Ochmann, Hussain, Von~Ahnen, Cordones,
  Hong, Lee, Ma, Adamczyk, Kim, Schoenlein, \latin{et~al.}
  others]{ochmann2018uv}
Ochmann,~M.; Hussain,~A.; Von~Ahnen,~I.; Cordones,~A.~A.; Hong,~K.; Lee,~J.~H.;
  Ma,~R.; Adamczyk,~K.; Kim,~T.~K.; Schoenlein,~R.~W. \latin{et~al.}
  UV-photochemistry of the disulfide bond: Evolution of early photoproducts
  from picosecond X-ray absorption spectroscopy at the sulfur K-Edge. \emph{J.
  Am. Chem. Soc.} \textbf{2018}, \emph{140}, 6554--6561\relax
\mciteBstWouldAddEndPuncttrue
\mciteSetBstMidEndSepPunct{\mcitedefaultmidpunct}
{\mcitedefaultendpunct}{\mcitedefaultseppunct}\relax
\EndOfBibitem
\bibitem[DeBeer~George \latin{et~al.}(2005)DeBeer~George, Brant, and
  Solomon]{debeer2005metal}
DeBeer~George,~S.; Brant,~P.; Solomon,~E.~I. Metal and ligand K-Edge XAS of
  organotitanium complexes: Metal 4p and 3d contributions to pre-edge intensity
  and their contributions to bonding. \emph{J. Am. Chem. Soc.} \textbf{2005},
  \emph{127}, 667--674\relax
\mciteBstWouldAddEndPuncttrue
\mciteSetBstMidEndSepPunct{\mcitedefaultmidpunct}
{\mcitedefaultendpunct}{\mcitedefaultseppunct}\relax
\EndOfBibitem
\bibitem[McKeown \latin{et~al.}(2011)McKeown, Gan, Pegg, Stolte, and
  Demchenko]{mckeown2011x}
McKeown,~D.~A.; Gan,~H.; Pegg,~I.~L.; Stolte,~W.~C.; Demchenko,~I. X-ray
  absorption studies of chlorine valence and local environments in borosilicate
  waste glasses. \emph{J. Nuc. Mat.} \textbf{2011}, \emph{408}, 236--245\relax
\mciteBstWouldAddEndPuncttrue
\mciteSetBstMidEndSepPunct{\mcitedefaultmidpunct}
{\mcitedefaultendpunct}{\mcitedefaultseppunct}\relax
\EndOfBibitem
\bibitem[Shadle \latin{et~al.}(1995)Shadle, Hedman, Hodgson, and
  Solomon]{shadle1995ligand}
Shadle,~S.~E.; Hedman,~B.; Hodgson,~K.~O.; Solomon,~E.~I. Ligand K-edge X-ray
  absorption spectroscopic studies: metal-ligand covalency in a series of
  transition metal tetrachlorides. \emph{J. Am. Chem. Soc.} \textbf{1995},
  \emph{117}, 2259--2272\relax
\mciteBstWouldAddEndPuncttrue
\mciteSetBstMidEndSepPunct{\mcitedefaultmidpunct}
{\mcitedefaultendpunct}{\mcitedefaultseppunct}\relax
\EndOfBibitem
\bibitem[Thom and Head-Gordon(2009)Thom, and Head-Gordon]{thom2009hartree}
Thom,~A.~J.; Head-Gordon,~M. Hartree--Fock solutions as a quasidiabatic basis
  for nonorthogonal configuration interaction. \emph{J. Chem. Phys.}
  \textbf{2009}, \emph{131}, 124113\relax
\mciteBstWouldAddEndPuncttrue
\mciteSetBstMidEndSepPunct{\mcitedefaultmidpunct}
{\mcitedefaultendpunct}{\mcitedefaultseppunct}\relax
\EndOfBibitem
\bibitem[List \latin{et~al.}(2020)List, Melin, van Horn, and
  Saue]{list2020beyond}
List,~N.~H.; Melin,~T. R.~L.; van Horn,~M.; Saue,~T. Beyond the electric-dipole
  approximation in simulations of x-ray absorption spectroscopy: Lessons from
  relativistic theory. \emph{J. Chem. Phys.} \textbf{2020}, \emph{152},
  184110\relax
\mciteBstWouldAddEndPuncttrue
\mciteSetBstMidEndSepPunct{\mcitedefaultmidpunct}
{\mcitedefaultendpunct}{\mcitedefaultseppunct}\relax
\EndOfBibitem
\bibitem[Rees \latin{et~al.}(2016)Rees, Wandzilak, Maganas, Wurster,
  Hugenbruch, Kowalska, Pollock, Lima, Finkelstein, and
  DeBeer]{rees2016experimental}
Rees,~J.~A.; Wandzilak,~A.; Maganas,~D.; Wurster,~N.~I.; Hugenbruch,~S.;
  Kowalska,~J.~K.; Pollock,~C.~J.; Lima,~F.~A.; Finkelstein,~K.~D.; DeBeer,~S.
  Experimental and theoretical correlations between vanadium K-edge X-ray
  absorption and K$\beta$ emission spectra. \emph{J. Bio. Inor. Chem.}
  \textbf{2016}, \emph{21}, 793--805\relax
\mciteBstWouldAddEndPuncttrue
\mciteSetBstMidEndSepPunct{\mcitedefaultmidpunct}
{\mcitedefaultendpunct}{\mcitedefaultseppunct}\relax
\EndOfBibitem
\bibitem[Farges(2009)]{farges2009chromium}
Farges,~F. Chromium speciation in oxide-type compounds: application to
  minerals, gems, aqueous solutions and silicate glasses. \emph{Phys. Chem.
  Minerals} \textbf{2009}, \emph{36}, 463--481\relax
\mciteBstWouldAddEndPuncttrue
\mciteSetBstMidEndSepPunct{\mcitedefaultmidpunct}
{\mcitedefaultendpunct}{\mcitedefaultseppunct}\relax
\EndOfBibitem
\bibitem[Hall \latin{et~al.}(2014)Hall, Pollock, Bendix, Collins, Glatzel, and
  DeBeer]{hall2014valence}
Hall,~E.~R.; Pollock,~C.~J.; Bendix,~J.; Collins,~T.~J.; Glatzel,~P.;
  DeBeer,~S. Valence-to-core-detected X-ray absorption spectroscopy: Targeting
  ligand selectivity. \emph{J. Am. Chem. Soc.} \textbf{2014}, \emph{136},
  10076--10084\relax
\mciteBstWouldAddEndPuncttrue
\mciteSetBstMidEndSepPunct{\mcitedefaultmidpunct}
{\mcitedefaultendpunct}{\mcitedefaultseppunct}\relax
\EndOfBibitem
\bibitem[Lancaster \latin{et~al.}(2011)Lancaster, Finkelstein, and
  DeBeer]{lancaster2011kbeta}
Lancaster,~K.~M.; Finkelstein,~K.~D.; DeBeer,~S. K$\beta$ X-ray emission
  spectroscopy offers unique chemical bonding insights: revisiting the
  electronic structure of ferrocene. \emph{Inorg. Chem.} \textbf{2011},
  \emph{50}, 6767--6774\relax
\mciteBstWouldAddEndPuncttrue
\mciteSetBstMidEndSepPunct{\mcitedefaultmidpunct}
{\mcitedefaultendpunct}{\mcitedefaultseppunct}\relax
\EndOfBibitem
\bibitem[Liu \latin{et~al.}(2011)Liu, Borg, Testemale, Etschmann, Hazemann, and
  Brugger]{liu2011speciation}
Liu,~W.; Borg,~S.~J.; Testemale,~D.; Etschmann,~B.; Hazemann,~J.-L.;
  Brugger,~J. Speciation and thermodynamic properties for cobalt chloride
  complexes in hydrothermal fluids at 35--440 C and 600 bar: an in-situ XAS
  study. \emph{Geochim. Cosmochim. Acta .} \textbf{2011}, \emph{75},
  1227--1248\relax
\mciteBstWouldAddEndPuncttrue
\mciteSetBstMidEndSepPunct{\mcitedefaultmidpunct}
{\mcitedefaultendpunct}{\mcitedefaultseppunct}\relax
\EndOfBibitem
\bibitem[DiMucci \latin{et~al.}(2019)DiMucci, Lukens, Chatterjee, Carsch,
  Titus, Lee, Nordlund, Betley, MacMillan, and Lancaster]{dimucci2019myth}
DiMucci,~I.~M.; Lukens,~J.~T.; Chatterjee,~S.; Carsch,~K.~M.; Titus,~C.~J.;
  Lee,~S.~J.; Nordlund,~D.; Betley,~T.~A.; MacMillan,~S.~N.; Lancaster,~K.~M.
  The myth of d$^8$ copper (III). \emph{J. Ame. Chem. Soc.} \textbf{2019},
  \emph{141}, 18508--18520\relax
\mciteBstWouldAddEndPuncttrue
\mciteSetBstMidEndSepPunct{\mcitedefaultmidpunct}
{\mcitedefaultendpunct}{\mcitedefaultseppunct}\relax
\EndOfBibitem
\bibitem[Balabanov and Peterson(2005)Balabanov, and
  Peterson]{balabanov2005systematically}
Balabanov,~N.~B.; Peterson,~K.~A. Systematically convergent basis sets for
  transition metals. I. All-electron correlation consistent basis sets for the
  3d elements Sc--Zn. \emph{J. Chem. Phys.} \textbf{2005}, \emph{123},
  064107\relax
\mciteBstWouldAddEndPuncttrue
\mciteSetBstMidEndSepPunct{\mcitedefaultmidpunct}
{\mcitedefaultendpunct}{\mcitedefaultseppunct}\relax
\EndOfBibitem
\bibitem[Yamamoto(2008)]{yamamoto2008assignment}
Yamamoto,~T. Assignment of pre-edge peaks in K-edge x-ray absorption spectra of
  3d transition metal compounds: electric dipole or quadrupole? \emph{X-Ray
  Spectrom.} \textbf{2008}, \emph{37}, 572--584\relax
\mciteBstWouldAddEndPuncttrue
\mciteSetBstMidEndSepPunct{\mcitedefaultmidpunct}
{\mcitedefaultendpunct}{\mcitedefaultseppunct}\relax
\EndOfBibitem
\bibitem[Southworth \latin{et~al.}(2019)Southworth, Dunford, Ray, Kanter,
  Doumy, March, Ho, Kr\"assig, Gao, Lehmann, Pic\'on, Young, Walko, and
  Cheng]{Southworth2019rel}
Southworth,~S.~H.; Dunford,~R.~W.; Ray,~D.; Kanter,~E.~P.; Doumy,~G.;
  March,~A.~M.; Ho,~P.~J.; Kr\"assig,~B.; Gao,~Y.; Lehmann,~C.~S.
  \latin{et~al.}  Observing pre-edge $K$-shell resonances in Kr, Xe, and
  ${\mathrm{XeF}}_{2}$. \emph{Phys. Rev. A} \textbf{2019}, \emph{100},
  022507\relax
\mciteBstWouldAddEndPuncttrue
\mciteSetBstMidEndSepPunct{\mcitedefaultmidpunct}
{\mcitedefaultendpunct}{\mcitedefaultseppunct}\relax
\EndOfBibitem
\bibitem[Breit(1932)]{breit1932dirac}
Breit,~G. Dirac's equation and the spin-spin interactions of two electrons.
  \emph{Phys. Rev.} \textbf{1932}, \emph{39}, 616--624\relax
\mciteBstWouldAddEndPuncttrue
\mciteSetBstMidEndSepPunct{\mcitedefaultmidpunct}
{\mcitedefaultendpunct}{\mcitedefaultseppunct}\relax
\EndOfBibitem
\bibitem[Kozio{\l} and Aucar(2018)Kozio{\l}, and Aucar]{koziol2018qed}
Kozio{\l},~K.; Aucar,~G.~A. QED effects on individual atomic orbital energies.
  \emph{J. Chem. Phys.} \textbf{2018}, \emph{148}, 134101\relax
\mciteBstWouldAddEndPuncttrue
\mciteSetBstMidEndSepPunct{\mcitedefaultmidpunct}
{\mcitedefaultendpunct}{\mcitedefaultseppunct}\relax
\EndOfBibitem
\bibitem[Szabo and Ostlund(1996)Szabo, and Ostlund]{szabo2012modern}
Szabo,~A.; Ostlund,~N.~S. \emph{{Modern Quantum Chemistry: Introduction to
  Advanced Electronic Structure Theory}}; Dover Publications, Inc.: Mineola,
  New York, 1996; pp 286--296\relax
\mciteBstWouldAddEndPuncttrue
\mciteSetBstMidEndSepPunct{\mcitedefaultmidpunct}
{\mcitedefaultendpunct}{\mcitedefaultseppunct}\relax
\EndOfBibitem
\bibitem[Zheng \latin{et~al.}(2022)Zheng, Zhang, Jin, Southworth, and
  Cheng]{zheng2022benchmark}
Zheng,~X.; Zhang,~C.; Jin,~Z.; Southworth,~S.~H.; Cheng,~L. Benchmark
  Relativistic Delta-Coupled-Cluster Calculations of K-Edge Core-Ionization
  Energies for Third-Row Elements. \emph{Phys. Chem. Chem. Phys.}
  \textbf{2022}, \emph{Submitted}\relax
\mciteBstWouldAddEndPuncttrue
\mciteSetBstMidEndSepPunct{\mcitedefaultmidpunct}
{\mcitedefaultendpunct}{\mcitedefaultseppunct}\relax
\EndOfBibitem
\bibitem[Wen and Hitchcock(1993)Wen, and Hitchcock]{wen1993inner}
Wen,~A.; Hitchcock,~A. Inner shell spectroscopy of
  \ce{($\eta$^5-C5H5)2TiCl2},\ce{($\eta$^5-C5H5) TiCl3}, and \ce{TiCl4}.
  \emph{Can. J. Chem.} \textbf{1993}, \emph{71}, 1632--1644\relax
\mciteBstWouldAddEndPuncttrue
\mciteSetBstMidEndSepPunct{\mcitedefaultmidpunct}
{\mcitedefaultendpunct}{\mcitedefaultseppunct}\relax
\EndOfBibitem
\bibitem[Mitzner \latin{et~al.}(2013)Mitzner, Rehanek, Kern, Gul, Hattne,
  Taguchi, Alonso-Mori, Tran, Weniger, Schr\"oder, \latin{et~al.}
  others]{mitzner2013edge}
Mitzner,~R.; Rehanek,~J.; Kern,~J.; Gul,~S.; Hattne,~J.; Taguchi,~T.;
  Alonso-Mori,~R.; Tran,~R.; Weniger,~C.; Schr\"oder,~H. \latin{et~al.}  L-edge
  x-ray absorption spectroscopy of dilute systems relevant to metalloproteins
  using an x-ray free-electron laser. \emph{J. Phys. Chem. Lett.}
  \textbf{2013}, \emph{4}, 3641--3647\relax
\mciteBstWouldAddEndPuncttrue
\mciteSetBstMidEndSepPunct{\mcitedefaultmidpunct}
{\mcitedefaultendpunct}{\mcitedefaultseppunct}\relax
\EndOfBibitem
\bibitem[Hocking \latin{et~al.}(2006)Hocking, Wasinger, de~Groot, Hodgson,
  Hedman, and Solomon]{hocking2006fe}
Hocking,~R.~K.; Wasinger,~E.~C.; de~Groot,~F.~M.; Hodgson,~K.~O.; Hedman,~B.;
  Solomon,~E.~I. Fe L-edge XAS studies of \ce{K4[Fe(CN)6]} and
  \ce{K3[Fe(CN)6]}: a direct probe of back-bonding. \emph{J. Am. Chem. Soc.}
  \textbf{2006}, \emph{128}, 10442--10451\relax
\mciteBstWouldAddEndPuncttrue
\mciteSetBstMidEndSepPunct{\mcitedefaultmidpunct}
{\mcitedefaultendpunct}{\mcitedefaultseppunct}\relax
\EndOfBibitem
\bibitem[Wen \latin{et~al.}(1992)Wen, R{\"u}hl, and Hitchcock]{wen1992inner}
Wen,~A.; R{\"u}hl,~E.; Hitchcock,~A. Inner-shell excitation of organoiron
  compounds by electron impact. \emph{Organometallics} \textbf{1992},
  \emph{11}, 2559--2569\relax
\mciteBstWouldAddEndPuncttrue
\mciteSetBstMidEndSepPunct{\mcitedefaultmidpunct}
{\mcitedefaultendpunct}{\mcitedefaultseppunct}\relax
\EndOfBibitem
\bibitem[Epifanovsky \latin{et~al.}(2021)Epifanovsky, \latin{et~al.}
  others]{epifanovsky2021software}
Epifanovsky,~E., \latin{et~al.}  Software for the frontiers of quantum
  chemistry: An overview of developments in the Q-Chem 5 package. \emph{J.
  Chem. Phys.} \textbf{2021}, \emph{155}, 084801\relax
\mciteBstWouldAddEndPuncttrue
\mciteSetBstMidEndSepPunct{\mcitedefaultmidpunct}
{\mcitedefaultendpunct}{\mcitedefaultseppunct}\relax
\EndOfBibitem
\bibitem[Perdew \latin{et~al.}(1982)Perdew, Parr, Levy, and
  Balduz~Jr]{perdew1982density}
Perdew,~J.~P.; Parr,~R.~G.; Levy,~M.; Balduz~Jr,~J.~L. {Density-functional
  theory for fractional particle number: derivative discontinuities of the
  energy}. \emph{Phys. Rev. Lett.} \textbf{1982}, \emph{49}, 1691--1694\relax
\mciteBstWouldAddEndPuncttrue
\mciteSetBstMidEndSepPunct{\mcitedefaultmidpunct}
{\mcitedefaultendpunct}{\mcitedefaultseppunct}\relax
\EndOfBibitem
\bibitem[Hait and Head-Gordon(2018)Hait, and
  Head-Gordon]{hait2018delocalization}
Hait,~D.; Head-Gordon,~M. Delocalization errors in density functional theory
  are essentially quadratic in fractional occupation number. \emph{J. Phys.
  Chem. Lett.} \textbf{2018}, \emph{9}, 6280--6288\relax
\mciteBstWouldAddEndPuncttrue
\mciteSetBstMidEndSepPunct{\mcitedefaultmidpunct}
{\mcitedefaultendpunct}{\mcitedefaultseppunct}\relax
\EndOfBibitem
\bibitem[Sun \latin{et~al.}(2020)Sun, Zhang, Banerjee, Bao, Barbry, Blunt,
  Bogdanov, Booth, Chen, Cui, \latin{et~al.} others]{sun2020recent}
Sun,~Q.; Zhang,~X.; Banerjee,~S.; Bao,~P.; Barbry,~M.; Blunt,~N.~S.;
  Bogdanov,~N.~A.; Booth,~G.~H.; Chen,~J.; Cui,~Z.-H. \latin{et~al.}  Recent
  developments in the PySCF program package. \emph{J. Chem. Phys.}
  \textbf{2020}, \emph{153}, 024109\relax
\mciteBstWouldAddEndPuncttrue
\mciteSetBstMidEndSepPunct{\mcitedefaultmidpunct}
{\mcitedefaultendpunct}{\mcitedefaultseppunct}\relax
\EndOfBibitem
\bibitem[Johnson~III(2015)]{johnson2015nist}
Johnson~III,~R.~D. {NIST Computational Chemistry Comparison and Benchmark
  Database, NIST Standard Reference Database Number 101, Release 18. October
  2016}. \emph{NIST} \textbf{2015}, \relax
\mciteBstWouldAddEndPunctfalse
\mciteSetBstMidEndSepPunct{\mcitedefaultmidpunct}
{}{\mcitedefaultseppunct}\relax
\EndOfBibitem
\bibitem[Groom \latin{et~al.}(2016)Groom, Bruno, Lightfoot, and
  Ward]{groom2016cambridge}
Groom,~C.~R.; Bruno,~I.~J.; Lightfoot,~M.~P.; Ward,~S.~C. The Cambridge
  structural database. \emph{Acta Crystallogr. B: Struct. Sci., Cryst. Eng.
  Mater.} \textbf{2016}, \emph{72}, 171--179\relax
\mciteBstWouldAddEndPuncttrue
\mciteSetBstMidEndSepPunct{\mcitedefaultmidpunct}
{\mcitedefaultendpunct}{\mcitedefaultseppunct}\relax
\EndOfBibitem
\bibitem[Mardirossian and Head-Gordon(2016)Mardirossian, and
  Head-Gordon]{wB97MV}
Mardirossian,~N.; Head-Gordon,~M. {$\omega$B97M-V: A combinatorially optimized,
  range-separated hybrid, meta-GGA density functional with VV10 nonlocal
  correlation}. \emph{J. Chem. Phys.} \textbf{2016}, \emph{144}, 214110\relax
\mciteBstWouldAddEndPuncttrue
\mciteSetBstMidEndSepPunct{\mcitedefaultmidpunct}
{\mcitedefaultendpunct}{\mcitedefaultseppunct}\relax
\EndOfBibitem
\bibitem[McGinnety(1972)]{mcginnety1972cesium}
McGinnety,~J.~A. Cesium tetrachlorocuprate. Structure, crystal forces, and
  charge distribution. \emph{J. Am. Chem. Soc.} \textbf{1972}, \emph{94},
  8406--8413\relax
\mciteBstWouldAddEndPuncttrue
\mciteSetBstMidEndSepPunct{\mcitedefaultmidpunct}
{\mcitedefaultendpunct}{\mcitedefaultseppunct}\relax
\EndOfBibitem
\bibitem[Franck and Dymond(1926)Franck, and Dymond]{franck1926elementary}
Franck,~J.; Dymond,~E. Elementary processes of photochemical reactions.
  \emph{Trans. Far. Soc.} \textbf{1926}, \emph{21}, 536--542\relax
\mciteBstWouldAddEndPuncttrue
\mciteSetBstMidEndSepPunct{\mcitedefaultmidpunct}
{\mcitedefaultendpunct}{\mcitedefaultseppunct}\relax
\EndOfBibitem
\bibitem[Condon(1926)]{condon1926theory}
Condon,~E. A theory of intensity distribution in band systems. \emph{Phys.
  Rev.} \textbf{1926}, \emph{28}, 1182--1201\relax
\mciteBstWouldAddEndPuncttrue
\mciteSetBstMidEndSepPunct{\mcitedefaultmidpunct}
{\mcitedefaultendpunct}{\mcitedefaultseppunct}\relax
\EndOfBibitem
\bibitem[Kutzelnigg(1984)]{kutzelnigg1984basis}
Kutzelnigg,~W. Basis set expansion of the Dirac operator without variational
  collapse. \emph{Int. J. Quantum Chem.} \textbf{1984}, \emph{25},
  107--129\relax
\mciteBstWouldAddEndPuncttrue
\mciteSetBstMidEndSepPunct{\mcitedefaultmidpunct}
{\mcitedefaultendpunct}{\mcitedefaultseppunct}\relax
\EndOfBibitem
\bibitem[Smith \latin{et~al.}(2020)Smith, Burns, Simmonett, Parrish, Schieber,
  Galvelis, Kraus, Kruse, Di~Remigio, Alenaizan, \latin{et~al.}
  others]{smith2020psi4}
Smith,~D.~G.; Burns,~L.~A.; Simmonett,~A.~C.; Parrish,~R.~M.; Schieber,~M.~C.;
  Galvelis,~R.; Kraus,~P.; Kruse,~H.; Di~Remigio,~R.; Alenaizan,~A.
  \latin{et~al.}  PSI4 1.4: Open-source software for high-throughput quantum
  chemistry. \emph{J. Chem. Phys.} \textbf{2020}, \emph{152}, 184108\relax
\mciteBstWouldAddEndPuncttrue
\mciteSetBstMidEndSepPunct{\mcitedefaultmidpunct}
{\mcitedefaultendpunct}{\mcitedefaultseppunct}\relax
\EndOfBibitem
\bibitem[Carroll \latin{et~al.}(1987)Carroll, Ji, Maclaren, Thomas, and
  Saethre]{carroll1987relativistic}
Carroll,~T.~X.; Ji,~D.; Maclaren,~D.~C.; Thomas,~T.~D.; Saethre,~L.~J.
  Relativistic corrections to reported sulfur 1s ionization energies. \emph{J.
  Electron Spectrosc. Relat. Phenom.} \textbf{1987}, \emph{42}, 281--284\relax
\mciteBstWouldAddEndPuncttrue
\mciteSetBstMidEndSepPunct{\mcitedefaultmidpunct}
{\mcitedefaultendpunct}{\mcitedefaultseppunct}\relax
\EndOfBibitem
\bibitem[Cavell and Sodhi(1987)Cavell, and Sodhi]{cavell1987effect}
Cavell,~R.; Sodhi,~R. The effect of relativistic curvature calibration
  corrections for a non-retarding hemispherical sector analyzer: revision of
  the absolute S1s binding energy of H2S and the KLL auger energy of PH3.
  \emph{J. Electron Spectrosc. Relat. Phenom.} \textbf{1987}, \emph{43},
  215--223\relax
\mciteBstWouldAddEndPuncttrue
\mciteSetBstMidEndSepPunct{\mcitedefaultmidpunct}
{\mcitedefaultendpunct}{\mcitedefaultseppunct}\relax
\EndOfBibitem
\bibitem[Bourne~Worster \latin{et~al.}(2021)Bourne~Worster, Feighan, and
  Manby]{bourne2021reliable}
Bourne~Worster,~S.; Feighan,~O.; Manby,~F.~R. Reliable transition properties
  from excited-state mean-field calculations. \emph{J. Chem. Phys.}
  \textbf{2021}, \emph{154}, 124106\relax
\mciteBstWouldAddEndPuncttrue
\mciteSetBstMidEndSepPunct{\mcitedefaultmidpunct}
{\mcitedefaultendpunct}{\mcitedefaultseppunct}\relax
\EndOfBibitem
\end{mcitethebibliography}
\end{document}